\documentclass[a4paper,11pt]{article}  
\usepackage[top=3.5cm,bottom=3.5cm,left=3cm,right=3cm]{geometry}  

\usepackage{amsmath}  
\usepackage{amssymb}  
\usepackage{cite}
\usepackage{orcidlink}  
\usepackage{authblk}

\allowdisplaybreaks

\usepackage{xcolor}  

\definecolor{mateoblue}{rgb}{0, 0.5, 0.5}  

\definecolor{improveblue2}{rgb}{0, 0, 0.5}  

\definecolor{rewrittenblue}{rgb}{0.1, 0.2, 0.5}  

\newcommand{\Ham}{\mathcal{H}}  
\newcommand{\Hamorder}[1]{\mathcal{H}^{(#1)}}  
\newcommand{\pphi}{p_\phi \,}  
\newcommand{\pphiN}[1]{{p_\phi}^{\!#1}}  
\newcommand{\ppsi}{p_\psi \,}  
\newcommand{\ppsiN}[1]{{p_\psi}^{\!#1}}  

\newcommand{\hod}[1]{\mathring{h}_{#1}} 
\newcommand{\hou}[1]{\mathring{h}^{#1}} 

\newcommand{\piou}[1]{\mathring{\pi}^{#1}} 

\newcommand{\delh}[1]{\delta h_{#1}}
\newcommand{\delpi}[1]{\delta \pi^{#1}}

\begin{document}

\title{\textbf{Hamiltonian formalism for gauge-invariant cosmological perturbations with multiple scalar fields}} 
\author{\textbf{Mateo Pascual}
\orcidlink{0000-0001-7706-3075}}
\affil{\normalsize \textit{Dept.\,of Physics \& Astronomy, Western University, N6A\,3K7 London ON, Canada} \\ mpascua@uwo.ca}

\date{\today}

\maketitle


\begin{abstract}
We generalise Langlois' Hamiltonian treatment of gauge-invariant linear cosmological perturbations \cite{Langlois:1994ec} to a cosmological setting with multiple scalar fields minimally coupled to gravity. 
We review the Hamilton–Jacobi–\textit{like} technique for a Hamiltonian system with first-class constraints.
With this technique, elucidating the gauge-invariant quantities of the system is reduced to solving for the generating function of the appropriate canonical transformation. 
We then apply it to the case with only two scalar fields, showcasing how the presence of more than one scalar field results in a coupled evolution of the different gauge-invariant scalar degrees of freedom.
Their coupled evolution may lead to new phenomenology that cannot be simply inferred by superimposing the results of the single-field case.
A simple tracking of the scalar degrees of freedom present in a system with an arbitrary number of scalar fields drives the conclusion that the results for the two-field case can be trivially extended to a multi-field scenario 
For completeness, we conclude by recasting explicitly the derivation for the tensor modes, which is nevertheless unaffected by the number of scalar fields in the system. 
The resulting gauge-invariant Hamiltonian formulation provides a solid foundation for a canonical quantisation of perturbations, necessary to explore the quantum nature of the early universe.
\end{abstract}

\section{Introduction}
Cosmological perturbation theory \cite{Bardeen:1980kt, Kodama:1984ziu, Mukhanov:1988, Mukhanov:1990me}
has become a cornerstone of modern cosmology, providing the framework to analyse the evolution of primordial inhomogeneities that seed the large-scale structure of the universe
(see \cite{Mukhanov:2005book, Peter:2013woa, Baumann:2022book} for pedagogical sources).
The connection between quantum fluctuations in the early universe and the nearly scale-invariant spectrum of density perturbations has been extensively studied within this framework, enabling precise predictions of cosmic microwave background anisotropies and the matter power spectrum observed in galaxy surveys 
\cite{Planck:2018vyg, SDSS:2005xqv}.
It has been applied to inflationary models 
\cite{Sasaki:1986hm, Langlois:2010xc, Martin:2013tda}
as well as alternative scenarios such as the matter-bounce and the ekpyrotic scenario 
\cite{Brandenberger:2009jq, Buchbinder:2007ad, Brandenberger:2012zb, Wilson-Ewing:2012lmx, Cai:2013kja}
, allowing for a comparative analysis of their distinct observational imprints such as their predicted primordial power spectrum, non-Gaussianity, or tensor modes 
\cite{Planck:2018jri, Agullo:2020cvg, Agullo:2017eyh}.

A Hamiltonian formulation of cosmological perturbations offers an alternative perspective to the conventional Lagrangian approach 
(see \cite{Baumann:2022book, Nandi:2015ogk} and references therein),
and is particularly advantageous in the context of quantum cosmology. 
The seminal work of Langlois 
\cite{Langlois:1994ec} 
established a Hamiltonian formalism for linear cosmological perturbations in the case of a single scalar field, laying the groundwork for a constraint-based approach to gauge-invariant perturbations. 
The formalism presented there prescribes a systematic treatment of gauge invariance and a transparent identification of the canonical structure, providing a natural setting for the canonical quantisation of cosmological perturbations – a key step in investigating the quantum nature of the early universe.

Furthermore, the early universe serves as an exceptional test-bed to probe the validity of different candidate quantum gravity theories.
The Hamiltonian formulation is well suited for studying quantum gravity effects on perturbations, making it a natural tool for assessing the robustness of early-universe models within quantum gravity frameworks \cite{Anderegg:1994xq}.
Cosmological scenarios based on Hamiltonian methods have found particular utility in approaches such as loop quantum cosmology, where the quantisation of spacetime itself plays a central role and the initial singularity is resolved by a cosmic bounce 
\cite{Agullo:2016tjh, Ashtekar:2011ni, Bojowald:2008zzb}.

The purpose of this work is to generalise the existing Hamiltonian formalism laid out by Langlois \cite{Langlois:1994ec} to a system with multiple scalar fields comprising its matter sector, providing a clear prescription for extracting the physical modes of cosmological perturbations.
The coupled evolution of the perturbations may lead to novel phenomenology that cannot simply be inferred by superimposing single-field results (see for example \cite{Frion:2025}).
For pedagogical completeness, we will explicitly carry out the same steps undergone in Langlois’ derivation applied to a system of two scalar fields, extracting the physical, gauge-invariant degrees of freedom of the system and their equations of motion. We then extend this derivation to an arbitrary number of scalar fields minimally coupled to gravity.  This extension is mostly straight-forward, since we are only concerned with linear perturbations and will not consider terms past second-order in perturbations (see \cite{Domenech:2017ems} for second-order perturbations and \cite{Nandi:2015ogk} for expansions of the Hamiltonian to further orders), and will only involve a careful tracking of the available scalar degrees of freedom in the system.


The paper is organised as follows. 
In Section 2, we review the Hamiltonian formulation of general relativity by means of the ADM decomposition and emphasise the role of first-class constraints in enforcing gauge invariance.
Section 3 is devoted to the description of the homogeneous and isotropic Friedmann-Lemaître-Robertson-Walker (FLRW)  background, where the dynamics is reduced to a few key variables. 
In Section 4, we introduce linear perturbations around the FLRW background and detail the split into background and inhomogeneous parts. 
In Section 5 we review the Hamilton–Jacobi–\textit{like} method employed to extract the gauge-invariant degrees of freedom by means of canonical transformations, setting the stage for a transparent treatment of the physical modes. 
The scalar sector is analysed in detail in Section 6, where we derive the scalar gauge-invariant Hamiltonian and the equations of motion for the scalar gauge-invariant variables. 
In Section 7, we extend the analysis of the scalar sector to incorporate multiple scalar fields, discussing the new subtleties that arise from the additional degrees of freedom when the Hamiltonian is expanded to second-order in perturbations. 
Finally, in Section 8 we treat the tensor sector explicitly for pedagogical completeness even though it is not different to the case with only one scalar field. We highlight the explicit derivation of the tensor modes and draw a parallelism with the scalar sector. 
We conclude in Section 9 with a brief discussion of our results.

\section{Hamiltonian formulation of General Relativity}
Let us begin by reviewing the Hamiltonian formulation of general relativity (see e.g. \cite{wald_general_1984}) considering a matter sector composed of two scalar fields.
Although this is standard textbook lore, it will serve us to fix the notation employed and keep the paper self-contained.
In a 4-dimensional manifold endowed with a metric $g_{\mu\nu}$, the dynamics of the gravitational field together with two scalar fields $\Phi$ and $\Psi$ minimally coupled to gravity is embodied in the action
\begin{equation} \label{Action}
    I = \int \mathrm{d}^4x \sqrt{-g} \left[ \frac{1}{2\kappa} R 
    - \frac{1}{2} g^{\mu\nu} \partial_\mu \Phi \partial_\nu \Phi - U(\Phi) 
    - \frac{1}{2} g^{\mu\nu} \partial_\mu \Psi \partial_\nu \Psi - V(\Psi) \right] \ ,
\end{equation}
where $\kappa = 8\pi G$ (and we set $c=1$), $R$ is the 4-Ricci scalar curvature, and $U(\Phi)$ and $V(\Psi)$ denote the arbitrary potentials for the scalar fields. To derive a Hamiltonian formulation from this action we employ the ADM formalism: a foliation of spacelike 3-surfaces is introduced via a time function $t$, and these surfaces are parametrised by coordinates $x^i$. The spacetime metric is then decomposed into the lapse $N(t,\mathbf{x})$, the shift $N^i(t, \mathbf{x})$, and the 3-metric $h_{ij}(t,\mathbf{x})$ as
\begin{equation}
    \mathrm{d}s^2 = -(N^2 - N_i N^i) \, \mathrm{d}t^2+ 2N_i \, \mathrm{d}x^i \mathrm{d}t + h_{ij} \, \mathrm{d}x^i \mathrm{d}x^j \ .
\end{equation}
Moreover, the 4-curvature is expressed in terms of the Ricci scalar on the spatial 3-surface $^{(3)}\!R$ and the extrinsic curvature tensor
\begin{equation}
    K_{ij} = \frac{1}{2N} \left[ -\dot{h}_{ij} + N_{i|j} + N_{j|i} \right] \ ,
\end{equation}
where a dot denotes differentiation with respect to $t$ and a vertical stroke preceding an index indicates the spatial covariant derivative compatible with $h_{ij}$. This decomposition reads
\begin{equation}
    R = {}^{(3)}\!R + K_{ij} K^{ij} - K^2 \ ,
\end{equation}
where $h^{ij}$ – the inverse of the 3-metric $h_{ij}$ given by $h_{ij} h^{jk} = \delta_i^k$ (i.e. the Kronecker delta)~– is used to raise the spatial indices of $K_{ij}$, and $K$ represents the trace $h^{ij} K_{ij}$.

The standard definitions for the canonical momenta conjugate to $h_{ij}$, $\Phi$, and $\Psi$ are
\begin{equation}
    \begin{gathered}
        \pi^{ij} := \frac{\delta I}{\delta \dot{h}_{ij}} 
        = \frac{\sqrt{h}}{2\kappa} \left( K h^{ij} - K^{ij} \right) \ ,
        \\[6pt]
        \pi_\Phi := \frac{\delta I}{\delta \dot{\Phi}} \ , \qquad  
        \pi_\Psi := \frac{\delta I}{\delta \dot{\Psi}} \ ,
        \\[4pt]
    \end{gathered}
\end{equation}
with $h = |h_{ij}|$ being the determinant of the 3-metric $h_{ij}$. These allow us to recast the action as 
\begin{equation}
    I = \int \mathrm{d}t \int \mathrm{d}^3x \left[\pi^{ij}\dot{h}_{ij} + \pi_\Phi \dot{\Phi} + \pi_\Psi \dot{\Psi} -N\mathcal{H} - N^i\mathcal{H}_i \right] \ ,
\end{equation}
where
\begin{equation}
    \begin{split} \label{Hamiltonian-density}
        \mathcal{H} = \frac{2\kappa}{\sqrt{h}} \left(\pi^{ij}\pi_{ij} - \frac{1}{2} \pi^2 \right) 
        - \frac{\sqrt{h}}{2\kappa} \, {}^{(3)}\!R
        &+ \frac{1}{2\sqrt{h}} (\pi_\Phi)^2 
        + \sqrt{h}U(\Phi) 
        + \frac{\sqrt{h}}{2} \partial_i\Phi\partial^i\Phi 
        \\
        &+ \frac{1}{2\sqrt{h}} (\pi_\Psi)^2 
        + \sqrt{h}V(\Psi) 
        + \frac{\sqrt{h}}{2} \partial_i\Psi\partial^i\Psi \ ,
    \end{split}
\end{equation}

\begin{equation} \label{Hamiltonian-momentum-density}
    \mathcal{H}_i = - 2 \partial_k \big( h_{ij} \pi^{jk} \big)
    + \pi^{lm}\partial_i h_{lm} 
    + \pi_\Phi \partial_i \Phi
    + \pi_\Psi \partial_i \Psi \ ,
\end{equation}
with $\pi$ denoting the trace $h_{ij}\pi^{ij}$.
The vanishing of the variation of the action with respect to the lapse $N$ and the shift $N^i$ gives the scalar constraint $\mathcal{H} = 0$ and the momentum constraint $\mathcal{H}_i = 0$.

By employing the ADM formalism we have been able to reformulate the original Lagrangian description in \eqref{Action} in a Hamiltonian framework, which can be summarised as follows. The phase space is spanned by the canonically conjugate fields 
$$\Gamma = \{ h_{ij},  \pi^{ij},  \Phi,  \pi_\Phi,  \Psi, \pi_\Psi \} \ , $$ satisfying the Poisson brackets
\begin{equation}
    \begin{split}
        \left\{ h_{ij}(\mathbf{x}), \pi^{kl}(\mathbf{y}) \right\} =& \ \delta_i^{(k} \delta_j^{l)} \delta^{(3)}(\mathbf{x}-\mathbf{y}) \ ,
        \\[6pt]
        \Big\{ \Phi(\mathbf{x}), \pi_\Phi(\mathbf{y}) \Big\} =& \ \delta^{(3)}(\mathbf{x}-\mathbf{y}) \ ,
        \\[6pt]
        \Big\{ \Psi(\mathbf{x}), \pi_\Psi(\mathbf{y}) \Big\} =& \ \delta^{(3)}(\mathbf{x}-\mathbf{y}) \ ,
    \end{split}
\end{equation}
where $\delta_i^{(k} \delta_j^{l)} = \frac{1}{2} ( \delta_i^{k} \delta_j^{l} + \delta_i^{l} \delta_j^{k} )$ is the symmetrised Kronecker delta and $\delta^{(3)}(\mathbf{x})$ is the Dirac delta.
The physical sector of the phase space is confined to the region of the phase space that satisfies the scalar (or energy) constraint
\begin{equation}
    \Ham ( h_{ij},  \pi^{ij},  \Phi,  \pi_\Phi,  \Psi, \pi_\Psi ) = 0
\end{equation}
and the vector (or momentum) constraint
\begin{equation}
    \Ham_i ( h_{ij},  \pi^{ij},  \Phi,  \pi_\Phi,  \Psi, \pi_\Psi ) = 0 \ .
\end{equation}
The Poisson brackets between these constraints yield only linear combinations of the constraints themselves such that, in Dirac's terminology, they are said to be \textit{first class} constraints \cite{dirac_lectures_1964}, a property that underscores the gauge invariance of the system under spacetime diffeomorphisms.

Moreover, the Hamiltonian of the system takes the form
\begin{equation} \label{Hamiltonian}
    H[ h_{ij},  \pi^{ij},  \Phi,  \pi_\Phi,  \Psi, \pi_\Psi ] = \int \mathrm{d}^3x \left(N \Ham + N^i \Ham_i \right) \ ,
\end{equation}
where the lapse and shift are just arbitrary functions of the phase space variables.
The evolution in time of any function of the elementary conjugate variables is generated by its Poisson bracket with the Hamiltonian as in
\begin{equation}
\frac{\mathrm{d}}{\mathrm{d}t} F = \{F, H\} \ .
\end{equation}
In the case of canonical variables, these are simply the familiar Hamilton equations in field theory:
\begin{equation}
        \begin{split}
        \dot{h}_{ij} = \frac{\delta H}{\delpi{ij}}
        \ , \qquad &
        \dot{\pi}^{ij} = -\frac{\delta H}{\delh{ij}}
        \ ,
        \\[6pt]
        \dot{\Phi} = \frac{\delta H}{\delta \pi_\Phi}
        \ , \qquad &
        \dot{\pi}_\Phi = -\frac{\delta H}{\delta \Phi}
        \ ,
        \\[6pt]
        \dot{\Psi} = \frac{\delta H}{\delta \pi_\Psi}
        \ , \qquad &
        \dot{\pi}_\Psi = -\frac{\delta H}{\delta \Psi} \ . 
        \end{split}
\end{equation}
It is important to note that in \eqref{Hamiltonian} the Hamiltonian is formulated exclusively as a combination of constraints. This contrasts with many conventional theories, such as electromagnetism, where the Hamiltonian consists of both constraint terms and a distinct dynamical component. As will be discussed in the following section, in the linearised regime the Hamiltonian associated with the perturbed variables ceases to be merely a combination of constraints imposed on these perturbations and acquires a dynamical component.

\section{FLRW background}
It is generally assumed in cosmology that the early universe is well described by a solution of Einstein’s equations that closely approximates FLRW geometry. Let us consider the Hamiltonian formulation for a spatially flat, homogeneous and isotropic FLRW universe prior to undertaking the linearisation procedure. In this context, the phase space fields in $\Gamma$ can be separated into homogeneous background quantities and small inhomogeneities as
\begin{equation} \label{Perturbation-split}
    \begin{split}
        \Phi = \phi(t) + \delta \phi(t, \mathbf{x})
        \ , \qquad &
        \pi_\Phi = \pphi(t) + \delta \pphi(t, \mathbf{x}) \ , 
        \\[6pt]
        \Psi = \psi(t) + \delta \psi(t, \mathbf{x})
        \ , \qquad &
        \pi_\Psi = \ppsi(t) + \delta \ppsi(t, \mathbf{x}) \ , 
        \\[6pt]
        h_{ij} = \hod{ij}(t) + \delh{ij}(t, \mathbf{x})
        \ , \qquad &
        \pi^{ij} = \piou{ij}(t) + \delpi{ij}(t, \mathbf{x}) \ ,
    \end{split}
\end{equation}
such that
$$\Gamma_0 = \big\{ \hod{ij}(t),  \piou{ij}(t),  \phi(t),  \pphi(t),  \psi(t), \ppsi(t) \big\}$$
is the phase space of the homogeneous, spatially-flat FLRW background, and 
$$\Gamma_1 = \big\{\delh{ij}(t,\mathbf{x}), \, \delpi{ij}(t,\mathbf{x}), \, \delta \phi(t,\mathbf{x}), \, \delta \pphi(t,\mathbf{x}), \, \delta \psi(t,\mathbf{x}), \, \\ \delta \ppsi(t,\mathbf{x}) \big\}$$
is the phase space of the perturbations, which are purely inhomogeneous.
In the following, the arguments of both background and perturbation variables are suppressed for simplicity.

The background spacetime metric is of the form 
\begin{equation}
\mathrm{d} s^2 = -N(t)^2 \mathrm{d} t^2 + a(t)^2 \delta_{ij} \mathrm{d} x^i \mathrm{d} x^j \ ,
\end{equation}
where $\delta_{ij}$ is the Euclidean metric. In this symmetry-reduced setting, the otherwise $6 \times \infty^3$ degrees of freedom of the 3-metric are reduced to a single one, just as for the momenta, such that they can be recasted as
\begin{equation}
\hod{ij} = a^2 \delta_{ij}
\qquad \text{and} \qquad
\piou{ij} = \frac{\pi_a}{6a}  \delta^{ij} \ ,
\end{equation}
where $a$ is the scale factor and $\pi_a$ is its conjugate momentum.
Hereon, spatial indices $i,j,...$ are raised and lowered by the background metric $\hod{ij}$ and its inverse $\hou{ij}$, related via $\hod{ij} \hou{jk} =\delta_i^k$.
The background flat-FLRW phase space is therefore reduced to the six-dimensional space spanned by
$$\Gamma_0 = \{ a,  \pi_a,  \phi,  \pphi,  \psi, \ppsi \} \ .$$

The background scalar constraint is given by
\begin{equation} \label{Background-scalar-constraint}
\Hamorder{0} =
-\frac{\kappa\pi_a^2}{12 a}
+\frac{\pphiN{2}}{2a^3}
+ a^3 U(\phi)
+\frac{\ppsiN{2}}{2a^3}
+ a^3 V(\psi) \ ,
\end{equation}
while the vector constraint vanishes identically since all its terms contain spatial derivatives, which vanish for FLRW background quantities, such that this constraint is trivially satisfied.
The background Hamiltonian then reads
\begin{equation} \label{Hamiltonian-0-notdensity}
    H^{(0)} = \int \mathrm{d}^3x \, N(t) \, \Hamorder{0} \ ,
\end{equation}
which is constrained to vanish
\footnote{If one were to consider a non-compact spatial manifold like $\mathbb{R}^3$, integrals of background homogeneous quantities over the whole spatial manifold may diverge. To avoid this spurious divergence spatial integrals may be restricted to a fiducial cell with an arbitrarily large comoving volume $\mathcal{V}_0$, and ultimately take the limit $\mathcal{V}_0 \rightarrow \infty$ at the end of the calculation. At the classical level, this infrared regulator is manifest only in intermediate expressions and does not appear in the resulting physics, such that it can be safely disregarded in any case. Nonetheless, it has been argued that, in the quantum theory, this regulator may in fact carry a physical significance that cannot be diregarded – see e.g. \cite{Mele:2021gzx, Mele:2022quh} and references therein.}. 
The non-vanishing Poisson brackets are
\begin{equation}
    \big\{ \phi,\pphi \big\} = \big\{ \psi,\ppsi \big\} = \big\{ a,\pi_a \big\} = 1 \ .
\end{equation}
and the equations of motion for the background, given by Hamilton's equations, are found to be
\begin{equation} \label{EOM-background-vars}
    \begin{split}
        \dot{a} &= \frac{\delta H^{(0)}}{\delta \pi_\alpha} 
        = -N \frac{\kappa \pi_a}{6a} \ , 
        \\[4pt]
        \dot{\pi}_a &= -\frac{\delta H^{(0)}}{\delta a} 
        = N \left( -\frac{\kappa \pi_a^2}{12a^2} 
        + \frac{3}{2} \frac{\pphiN{2}}{a^4}
        + 3 a^2 U(\phi)
        + \frac{3}{2} \frac{\ppsiN{2}}{a^4}
        + 3 a^2 V(\psi) \right) \ , 
        \\[4pt]
        \dot{\phi} &= \frac{\delta H^{(0)}}{\delta \pi_\phi} 
        = N \frac{\pphi}{a^3} \ , 
        \\[4pt]
        \dot{p}_\psi&= -\frac{\delta H^{(0)}}{\delta \phi} 
        = -N a^3 \frac{\delta U}{\delta \phi} \ , 
        \\[4pt]
        \dot{\psi} &= \frac{\delta H^{(0)}}{\delta \pi_\psi} 
        = N \frac{\ppsi}{a^3} \ , 
        \\[4pt]
        \dot{p}_\psi &= -\frac{\delta H^{(0)}}{\delta \psi} 
        = -N a^3 \frac{\delta V}{\delta \psi} \ .
    \end{split}
\end{equation}

\section{Linear perturbations}
Let us now consider the dynamics of the small linear perturbations over the background variables as introduced in \eqref{Perturbation-split}.
The perturbations $\delh{ij}$, $\delpi{ij}$, $\delta \phi$, $\delta \pphi$, $\delta \psi$ and $\delta \ppsi$ are taken to be canonically conjugate coordinates on the linearised phase space $\Gamma_1$, such that their non-vanishing Poisson brackets are given by
\begin{equation} \label{Poisson-delta-h-delta-pi}
    \begin{aligned}    
        \left\{ \delh{ij}(\mathbf{x}) , \delpi{kl}(\mathbf{y}) \right\} 
        &= \delta_i^{(k} \delta_j^{l)} \, \delta^{(3)}(\mathbf{x} - \mathbf{y}) \ ,
        \\[6pt]
        \Big\{ \delta \phi (\mathbf{x}), \delta \pphi (\mathbf{y}) \Big\} 
        &= \, \delta^{(3)}(\mathbf{x} - \mathbf{y}) \ ,
        \\[6pt]
        \Big\{ \delta \psi (\mathbf{x}), \delta \ppsi (\mathbf{y}) \Big\} 
        &= \, \delta^{(3)}(\mathbf{x} - \mathbf{y}) \ ,
    \end{aligned}
\end{equation}
while all other Poisson brackets vanish. The perturbation variables in $\Gamma_1$ must satisfy the linearised versions of the constraints given by substituting \eqref{Perturbation-split} into \eqref{Hamiltonian-density} and \eqref{Hamiltonian-momentum-density}. Furthermore, the dynamics of the perturbations will be embodied in the second-order terms in the expansion of the Hamiltonian in the perturbative quantities. In this sense, the linearisation procedure introduces a split between the vanishing constraints and the dynamical terms in the Hamiltonian.

Proceeding with the linearisation of the constraints around the flat-FLRW background discussed in the previous section, we obtain
\begin{equation} \label{Linear-constraint-Ham}
    \begin{split}
    \Ham^{(1)} = \;
    &\; \frac{\kappa}{a^3} \left(
    -\frac{1}{72} a^2 \pi_a^2 \hou{ij} \delh{ij} - \frac{1}{3} a \pi_a \hod{ij} \delpi{ij} \right)
    - \frac{a^3}{2 \kappa} \left( \hou{il} \hou{jk} - \hou{ik} \hou{jl} \right) \partial_j \partial_l \delh{ik}
    \\[4pt]
    & + \frac{\pphi}{a^3} \delta \pphi - \frac{1}{4} \frac{\pphiN{2}}{a^3} \hou{ij} \delh{ij} + a^3 \frac{\partial U}{\partial \phi} \delta \phi + \frac{1}{2} a^3 U(\phi) \hou{ij} \delh{ij}
    \\[4pt]
    & + \frac{\ppsi}{a^3} \delta \ppsi - \frac{1}{4} \frac{\ppsiN{2}}{a^3} \hou{ij} \delh{ij} + a^3 \frac{\partial V}{\partial \psi} \delta \psi + \frac{1}{2} a^3 V(\psi) \hou{ij} \delh{ij}
    \end{split}
\end{equation}
for the linearised scalar constraint, and 
\begin{equation} \label{Linear-constraint-Ham_i}
\Ham_i^{(1)} = 
-\frac{1}{3} a \pi_a \hou{jk} \partial_k \delh{ij} 
- 2 \hod{ij} \partial_k \delpi{jk} 
+ \frac{1}{6} a \pi_a \hou{jk} \partial_i \delh{jk} 
+ \pphi \partial_i \delta \phi 
+ \ppsi \partial_i \delta \psi
\end{equation}  
for the linearised vector constraint.
The second-order part of the Hamiltonian, which governs the dynamics of the perturbations, is given by
\begin{equation} \label{Ham-second-order-ito-densities}
H^{(2)} [\delh{ij}, \delpi{ij}, \delta \phi, \delta \pphi, \delta \psi, \delta \ppsi] = \int \mathrm{d}^3x \left(N \Hamorder{2} + N^i \Hamorder{2}_i \right) \ ,
\end{equation}
where $\Hamorder{2}$ and $\Hamorder{2}_i$ denote the second-order contributions to the scalar and momentum constraints, respectively.

At this stage, it is convenient to decompose the gravitational perturbations according to the symmetries of the background spatial metric $\hod{ij}$, namely the \textit{scalar-vector-tensor} decomposition of metric perturbations. This is commonly performed in Fourier space, with Fourier quantities defined by the formula
\begin{equation}
    \tilde{\delh{ij}}(\mathbf{k}) := \int \frac{\mathrm{d}^3 x}{(2\pi)^3} \mathrm{e}^{-i\mathbf{k} \cdot \mathbf{x}} \delh{ij}(\mathbf{x}) \ ,
\end{equation}
with analogous expressions for all other perturbation fields.
Hereon, the tilde of $\tilde{\delh{ij}}(\mathbf{k})$ and other perturbation fields in Fourier space is suppressed for simplicity, and it should be clear from its argument or the context whether a variable is valued in coordinate or Fourier space.

The $3 \times 3$ symmetric matrices $\delh{ij}(\mathbf{k})$ span a $(6 \times \infty^3)$-dimensional vector space. For each $\mathbf{k}$, this 6-dimensional space can be decomposed into three 2-dimensional subspaces, corresponding to the \textit{scalar}, \textit{vector} and \textit{tensor} components.
We obtain such a decomposition by writing $\delh{ij}(\mathbf{k})$ in a convenient basis $A_{ij}^{(m)}$ for $m = 1,...,6$ for this space, that is
\begin{equation} \label{A-basis}
    \begin{aligned}
        A_{ij}^{(1)} &= \hod{ij} \; , & \quad
        A_{ij}^{(2)} &= \left( \hat{k}_i \, \hat{k}_j - \frac{\hod{ij}}{3} \right) , \\[6pt]
        A_{ij}^{(3)} &= \frac{1}{\sqrt{2}} \left( \hat{k}_i \, \hat{x}_j + \hat{k}_j \, \hat{x}_i \right) , & \quad
        A_{ij}^{(4)} &= \frac{1}{\sqrt{2}} \left( \hat{k}_i \, \hat{y}_j + \hat{k}_j \, \hat{y}_i \right) , \\[6pt]
        A_{ij}^{(5)} &= \frac{1}{\sqrt{2}} \Big( \hat{x}_i \, \hat{y}_j + \hat{x}_j \, \hat{y}_i \Big) , & \quad
        A_{ij}^{(6)} &= \frac{1}{\sqrt{2}} \Big( \hat{x}_i \, \hat{y}_j - \hat{x}_j \, \hat{y}_i \Big) \ ,
    \end{aligned}
\end{equation}
where $\hat{\mathbf{k}}$ is the unit vector in the direction of $\mathbf{k}$, and $\hat{\mathbf{k}}$, $\hat{\mathbf{x}}$ and $\hat{\mathbf{y}}$ form an orthonormal set of unit vectors (with respect to $\hod{ij}$). These six matrices form an orthogonal basis with respect to the inner product
\begin{equation} \label{Inner-product-A}
    \left( \mathbf{A}^{\!(m)} , \mathbf{A}^{\!(n)} \right)
    := \mathrm{Trace}  \left( \mathbf{A}^{\!(m)} \mathbf{A}^{\!(n)} \right)
    = A_{ij}^{(m)} A_{kl}^{(n)} \hou{ik} \hou{jl}
    = A_{ij}^{(m)} A_{(n)}^{ij} \ ,
\end{equation}
such that
\begin{equation}
    \left( \mathbf{A}^{\!(m)} , \mathbf{A}^{\!(n)} \right)
    = \delta_{mn} \left( \mathbf{A}^{\!(m)} , \mathbf{A}^{\!(m)} \right) \ .
\end{equation}
Note that the position of the indices $m, n$ is not important and they will be placed wherever convenient.
We can now expand the metric perturbation field in this basis as
\begin{equation} \label{delta-h-expansion}
    \delh{ij}(\mathbf{k}) = \sum_{m=1}^6 \gamma_m(\mathbf{k}) A_{ij}^{(m)}(\mathbf{k}) \ .
\end{equation}
Having projected the metric perturbations onto this new basis, let us consider the $SO(2)$ subgroup of rotations around the direction $\hat{\mathbf{k}}$. Evidently, this group leaves $\hat{\mathbf{k}}$ invariant, but $\hat{\mathbf{x}}$ and $\hat{\mathbf{y}}$ are rotated by its action. Inspecting the definitions of the basis reveals that $A_{ij}^{(1)}$ and $A_{ij}^{(2)}$ are unaffected by the action of this group, while $A_{ij}^{(3)}$ and $A_{ij}^{(4)}$ transform as vectors, and $A_{ij}^{(5)}$ and $A_{ij}^{(6)}$ transform as two-covariant tensors. It is in this sense that $\gamma_m$ and $\pi_m$ are called \textit{scalar} modes for $m = 1, 2$, \textit{vector} modes for $m = 3, 4$, and \textit{tensor} modes for $m = 5, 6$. 
Note that although we have provided explicit expressions for the tensor and vector basis in \eqref{A-basis} for completeness, their explicit forms are not important – all that matters is that they satisfy the following relations:
\begin{equation}
    \begin{alignedat}{2}
        \hou{ij} A_{ij}^{(m)} &= 0 \ , \quad & m &= 3, \ldots, 6 \ ,
        \\[4pt]
        A_{ij}^{(m)} A^{ij}_{(n)} &= \delta_{m,n} \ , \qquad & m,n &= 3, \ldots, 6 \ , 
        \\[4pt]
        k^i k^j A_{ij}^{(m)} &= 0 \ , \quad & m &= 3, 4 \ , 
        \\[4pt]
        k^i A_{ij}^{(m)} &= 0 \ , \quad & m &= 5, 6 \ .
    \end{alignedat}
\end{equation}
Unlike the vector and tensor basis matrices which are of unit norm, the norm of the scalar basis with respect to \eqref{Inner-product-A} are given by
\footnote{In this work we adhere to the definition of $\mathbf{A}^{\!(1)}$ and $\mathbf{A}^{\!(2)}$ chosen by Langlois in \cite{Langlois:1994ec}. See, for example, \cite{Agullo:2017eyh} or \cite{Artigas:2021zdk} for an account where these matrices are chosen to be of unit norm, such that the whole basis is orthonormal.}
\begin{equation}
    \left( \mathbf{A}^{\!(1)} , \mathbf{A}^{\!(1)} \right) = A_{ij}^{(1)} A^{ij}_{(1)} = 3 
    \qquad \text{ and } \qquad
    \left( \mathbf{A}^{\!(2)} , \mathbf{A}^{\!(2)} \right) = A_{ij}^{(2)} A^{ij}_{(2)} = \frac{2}{3} \ .
\end{equation}
It is worth remarking that, given their orthogonality with respect to the inner product \eqref{Inner-product-A}, the matrices $\mathbf{A}^{(m)}$ have a background dependence of the form $a^2$ in order to compensate for the factor $a^{-2}$ in $\hou{ij}$ involved in raising the indices of the matrices to take their trace. This feature will be important later.

To replace the configuration coordinates $\delh{ij}$ by the $\gamma_m$ while preserving the canonical Poisson structure we will employ a generating function that relates the metric perturbation modes $\delh{ij}$ and $\delpi{ij}$ to the new variables $\gamma_m$ and $\pi_m$, such as
\begin{equation} \label{S-gamma}
    S_\gamma = -\delpi{ij} A_{ij}^{(m)} \gamma_m \ ,
\end{equation}
with integration over $\mathbf{k}$ being implied. The new canonical momenta $\pi_m$ are then obtained via
\begin{equation} \label{pi_m-definition}
    \pi_m = -\frac{\partial S_\gamma}{\partial \gamma_m} = A_{ij}^{(m)} \delpi{ij} \ .
\end{equation}
Furthermore, the new coordinates in this basis are found by inverting equation \eqref{delta-h-expansion} to be given by
\begin{equation} \label{gamma_m-definition}
    \gamma_m(\mathbf{k}) = \frac{A^{ij}_{(m)} \delh{ij}(\mathbf{k})}{\left( \mathbf{A}^{\!(m)} , \mathbf{A}^{\!(m)} \right)} \ .
\end{equation}
Inverting equation \eqref{pi_m-definition} we also find
\begin{equation} \label{delta-pi-expansion}
    \delpi{ij}(\mathbf{k}) = \sum_{m=1}^6 \frac{\pi_m(\mathbf{k}) A_{ij}^{(m)}(\mathbf{k})}{\left( \mathbf{A}^{\!(m)} , \mathbf{A}^{\!(m)} \right)} \ ,
\end{equation}
which will be useful later.

Using the above expressions and the Poisson brackets \eqref{Poisson-delta-h-delta-pi} we find the non-vanishing new Poisson brackets are
\begin{equation}
        \big\{ \gamma_m, \pi_n \big\} 
        = A^{ij}_{(m)} A_{kl}^{(n)} \times \left\{ \delh{ij}, \delpi{kl} \right\} = \, \delta_{mn} \ ,
\end{equation}
where a $\delta$-function in $\mathbf{k}$ is implied.
From \eqref{pi_m-definition} and \eqref{gamma_m-definition}, along with \eqref{A-basis}, we find that the scalar configuration variables are given by
\begin{equation} \label{Gamma-scalar}
    \gamma_1 = \frac{1}{3} \hou{ij} \delh{ij}  
    \qquad \text{and} \qquad
    \gamma_2 = \frac{3}{2} \left( \, \hat{k}^i \hat{k}^j - \frac{1}{3} \hou{ij} \right) \delh{ij} \ ,
\end{equation}
and are associated with their conjugate momenta given by
\begin{equation} \label{Pi-scalar}
    \pi_1 = \hod{ij} \delpi{ij}
    \qquad \text{and} \qquad
    \pi_2 = \left( {\hat{k}^i \hat{k}^j} - \frac{1}{3} \hod{ij} \right) \delpi{ij} \ .
\end{equation}
The use of the generating function guarantees that the new coordinates $\gamma_m$ and $\pi_m$ are indeed canonically conjugate. Thus, the Poisson brackets of the scalar variables are simply given by
\begin{equation}
    \{\gamma_1, \pi_1\} = \{\gamma_2, \pi_2\} = 1 \ ,
\end{equation}
\begin{equation}
    \{\gamma_1, \pi_2\} = \{\gamma_2, \pi_1\} = 0 \ ,
\end{equation}
where a $\delta$-function in $\mathbf{k}$ is implied. 

Furthermore, the linearised constraints can be decomposed just as well into their scalar and vector components (they do not have any tensor components), which act on their respective subspaces of the phase space. Regarding the scalar degrees of freedom, the linearised scalar constraint contains only scalar variables, such that mere substitution of the new variables given in \eqref{Gamma-scalar} and \eqref{Pi-scalar} into \eqref{Linear-constraint-Ham} yields
\begin{equation}
    \begin{aligned}
        E := 
        &- \frac{\kappa}{24} \frac{\pi_a^2}{a} \gamma_1 
        - \frac{\kappa}{3} \frac{\pi_a}{a^2} \pi_1 
        + \frac{a k^2}{3\kappa} \gamma_2 
        - \frac{a k^2}{\kappa} \gamma_1 
        \\
        &+ \frac{\pphi}{a^3} \delta \pphi 
        - \frac{3}{4} \frac{\pphiN{2}}{a^3} \gamma_1 
        + a^3 \frac{\partial U}{\partial \phi} \delta \phi 
        + \frac{3}{2} a^3 U(\phi) \gamma_1
        \\
        & + \frac{\ppsi}{a^3} \delta \ppsi 
        - \frac{3}{4} \frac{\ppsiN{2}}{a^3} \gamma_1 
        + a^3 \frac{\partial V}{\partial \psi} \delta \psi 
        + \frac{3}{2} a^3 V(\psi) \gamma_1 \equiv 0 \ ,
    \end{aligned}
\end{equation}
where $k^2:=k_i k_j \, \delta^{ij} = k_i k_j \, a^2 \hou{ij} = a^2 k_i k^i$ denotes the comoving wavenumber squared.
The~vector constraint actually contains one scalar and two vector components. By taking its divergence, all vector and tensor variables are eliminated due to the condition $k^i k^j A_{ij}^m = 0$, $m = 3,...,6$, satisfied by the matrix basis, and we are left with its scalar component as the constraint
\begin{equation}
    M :=
    \frac{1}{6} a \pi_a \gamma_1 
    - \frac{2}{9} a \pi_a \gamma_2 
    - \frac{2}{3} \pi_1
    - 2 \pi_2      + \pphi \delta \phi 
    + \ppsi \delta \psi \equiv 0 \ .
\end{equation}

\section{Gauge-invariant degrees of freedom} \label{sec:GI-DOFs}
In this section we will undergo the same procedure laid out by Goldberg et al. in \cite{goldberg_hamiltonian_1991} (as was undergone by Langlois in \cite{Langlois:1994ec}) to isolate the true physical degrees of freedom from the gauge ones in our system.
We begin by examining the scalar sector. The full scalar phase space is initially 8-dimensional, spanned by the variables $\gamma_1$, $\gamma_2$, $\delta \phi$, $\delta \psi$ and their conjugate momenta. Imposing the two first-class constraints, $E$ and $M$, restricts the system to a 6-dimensional subspace. 
Since the constraints are first-class, any point in this subspace can then undergo a gauge transformation and be mapped onto another point still within this constraint subspace. Having two such constraints implies that the gauge-orbits they generate are 2-dimensional, and the resulting physical space, given by the quotient of the constraint subspace by the gauge-orbits, is a 4-dimensional space. 

A similar counting applies to the vector and tensor modes.
The vector subspace of the phase space of linear perturbations is 4-dimensional, spanned by the two vector components of the metric perturbations. The two vector components of the linearised vector constraint \eqref{Linear-constraint-Ham_i} remove two degrees of freedom from the vector subspace, leaving a 2-dimensional constraint subspace. 
Furthermore, these two constraints on the vector subspace being first-class generate 2-dimensional gauge-orbits in the vector constraint-subspace, such that the resulting physical phase space is therefore null (zero-dimensional). There are no vector gauge-invariant degrees of freedom in the system.
Nevertheless, this would not be the case if matter of vector type were coupled to gravity, since in that case the vector subspace would be spanned by more than just the two metric perturbations' components.
The tensor subspace is similarly 4-dimensional from the start, spanned by $\gamma_5$, $\gamma_6$ and their conjugate momenta. In contrast, there are no constraints on the tensor subspace. The physical phase space of tensor modes is therefore 4-dimensional, and $\gamma_5$, $\gamma_6$ and their conjugate momenta are already gauge-invariant degrees of freedom.

To explicitly characterise this reduced, physical phase space, we will follow the procedure laid out by Goldberg et al. \cite{goldberg_hamiltonian_1991}, inspired by Hamilton-Jacobi theory (see, for example, \cite{goldstein_classical_1950} for further detail). Let us briefly recall a few basic notions from Hamilton–Jacobi theory and canonical transformations.
For a general $2n$-dimensional Hamiltonian system with Hamiltonian $H(q^i, p_i, t)$, where the $q^i$ and $p_i$ are canonically conjugate coordinates on the phase space with Poisson brackets $\{q^i, p_j\} = \delta^i_j$, a canonical transformation – represented by a \textit{generating function} $S(q^i, P_i, t)$ – preserves the structure of Hamilton's equations by preserving the Poisson bracket structure. The function $S$ generates the transformation because it relates the original coordinates and momenta to a new set $(Q^i, P_i)$ via
\begin{equation}
    p_i = \frac{\partial S}{\partial q^i} 
    \qquad \text{and} \qquad 
    Q^j = \frac{\partial S}{\partial P_j} \ ,
\end{equation}
with the new Hamiltonian given by
\begin{equation} \label{Hamiltonian-new-time-dependent}
    H_{\text{new}} = H + \frac{\partial S}{\partial t} \ .
\end{equation}
In fact, it may be that the Hamiltonian and the generating function have a time dependence implied by the explicit time dependence of some other \textit{external} variables $\lambda$ and $\pi_\lambda$ in the system, which in turn are governed by a separate external Hamiltonian of its own. In this case (which is the more relevant to this work), the expression for the new Hamiltonian generalises to
\begin{equation} \label{Hamiltonian-new-Poisson-bracket}
    H_{\text{new}} = H + \Big\{ S, H^\lambda \Big\}_{(\lambda)} \ , \\[4pt]
\end{equation}
where the Poisson bracket acts on the external variables.
The objective of Hamilton–Jacobi theory is to determine a canonical transformation that results in a vanishing new Hamiltonian. This transformation is embodied by a generating function \( S \) that satisfies the Hamilton–Jacobi equation
\begin{equation}
    \frac{\partial S}{\partial t} + H\left(q, p = \frac{\partial S}{\partial q}, t\right) = 0 \ .
\end{equation}
Once a \textit{complete} solution $S$ to this equation has been obtained — incorporating as many constants of integration as there are degrees of freedom — the problem of solving the equations of motion reduces to the more straightforward task of evaluating integrals.

The method introduced by Goldberg et al. in \cite{goldberg_hamiltonian_1991} is an alternative approach inspired by Hamilton–Jacobi theory, designed for a Hamiltonian system subject to a set of $m$ first-class constraints $C_\mu(q^i, p_i) = 0 \, (\mu = 1, \ldots, m)$. For a constrained Hamiltonian system like such, one can extract the $2(n-m)$ canonical coordinates that span the physical phase space by solving the Hamilton-Jacobi–\textit{like} equations
\begin{equation}
    C_\mu\left(q^i, \frac{\partial S}{\partial q^i}(q^i, P_a)\right) = 0 \ ,
\end{equation}
for the function $S = S(q^i, P_a)$, where $P_a$ are $(n-m)$ constants of integration with $(a = m+1, \ldots, n)$. 
The original coordinates are then related to the true, physical degrees of freedom $Q^a$ and $P_a$ by means of the singular canonical transformation generated by $S$ via the equations
\begin{equation}
    p_i = \frac{\partial S}{\partial q^i}
    \qquad \text{and} \qquad
    Q^a = \frac{\partial S}{\partial P_a} \ .
\end{equation}
Since there are less new coordinates than original coordinates, to invert these relations we must first introduce $m$ arbitrary parameters $\beta_\mu$, which represent arbitrary gauge functions, so that we may obtain $q^i = q^i(Q^a, P_a, \beta_\mu)$ and $p_i = p_i(Q^a, P_a, \beta_\mu)$. 
If the original Hamiltonian $H(q^i, p_i, t)$ is compatible with the first-class constraints $C_\mu$ (i.e., its Poisson bracket with any constraint vanishes weakly when the constraints are imposed), then the new Hamiltonian governing the true degrees of freedom is given by an expression like \eqref{Hamiltonian-new-time-dependent} or \eqref{Hamiltonian-new-Poisson-bracket}, replacing the variables $q^i$ and $p_i$ with their functions of $Q^a$ and $P_a$. Remarkably, the $\beta_\mu$ terms cancel out automatically in the new Hamiltonian such that we are able to simply ignore them.

Despite the similarities between standard Hamilton-Jacobi theory and the method we will employ, they are subtly different, both in their conception and objective.
In the standard approach, the Hamilton-Jacobi equation is built from the \textit{Hamiltonian} and is aimed at integrating the equations of motion, whereas in the method employed here, the Hamilton-Jacobi–\textit{like} equations are formulated using the \textit{constraints} with the purpose of reducing the phase space. 
Remarkably, the Hamilton-Jacobi-\textit{like} procedure we will be employing provides a clear and straightforward prescription for the singular canonical transformation – its generating function must satisfy the Hamilton–Jacobi version of the constraints – and directly yields the gauge-invariant degrees of freedom of the system along with its Hamiltonian.
Having reviewed it, we can now apply this method to our system of cosmological perturbations to extract the gauge-invariant quantities and find their equations of motion.

\section{Scalar sector} \label{sec:Scalar}
Since the scalar and tensor sectors of the gauge-invariant degrees of freedom are decoupled, we will study them separately – the former in this section and the latter in Section \ref{sec:Tensor}.
The scalar part of the linearised phase space is spanned by 
$$\Gamma_1^s = \big\{\gamma_1, \, \pi_1, \, \gamma_2, \, \pi_2, \, \delta \phi, \, \delta \pphi, \, \delta \psi, \, \delta \ppsi \big\} \ .$$
There are two scalar first-class constraints, namely $E$ and $M$. Applying the Hamilton–Jacobi--\-\textit{like} method recalled in section \ref{sec:GI-DOFs} to these constraints immediately yields the two equations
\begin{equation} \label{E-constraint}
    E\left(\gamma_\alpha, \pi_\beta = \frac{\partial S}{\partial \gamma_\beta}\right) = 0 \ ,
\end{equation}
\
\begin{equation} \label{M-constraint}
    M\left(\gamma_\alpha, \pi_\beta = \frac{\partial S}{\partial \gamma_\beta}\right) = 0, \ 
\end{equation}
where $\alpha, \beta = 1, 2, 3, 4$, and we temporarily repurpose 
the symbols $\gamma_3 := \delta \phi$, $\gamma_4 := \delta \psi$, $\pi_3 := \delta \pphi$ and $\pi_4 := \delta \ppsi$ for the analysis of the scalar sector that follows.
We adopt this labelling just as a convenient shorthand for the scalar phase space.
\footnote{We emphasise that $\gamma_\alpha$ with Greek indices represent the scalar sector of perturbations while $\gamma_m$ with Latin indices represent the decomposition of the metric perturbations.}
A solution $S$ for the system formed by \eqref{E-constraint} and \eqref{M-constraint} will provide a change of coordinates that would simplify the structure of the phase space. Since the system is linear, we may take an ansatz $S_{\scriptscriptstyle P}$ of quadratic order for the generating function $S$, such as
\begin{equation} \label{S_P-ansatz}
     S_{\scriptscriptstyle P} (\gamma_\alpha,P_\alpha) = \frac{1}{2} A_{\alpha \beta} \gamma_\alpha \gamma_\beta + B_\alpha \gamma_\alpha \ ,
\end{equation}
where the coefficients $A_{\alpha \beta}$ are symmetric and integration over $\mathbf{k}$ is implied. 
It then follows that
\begin{equation} \label{pi-alpha-action}
    \pi_\alpha = \frac{\partial S_{\scriptscriptstyle P}}{\partial \gamma_\alpha} = A_{\alpha \beta} \gamma_\beta + B_\alpha
\end{equation}
can be substituted back into \eqref{E-constraint} and \eqref{M-constraint} to get expressions in the form of a linear combination of $\gamma_\alpha$. Each of the coefficients of $\gamma_\alpha$ in $E$ and $M$ must vanish, as well as the sum of the constants, since the constraints must vanish identically. This creates a system of 5+5 equations for 10 $A_{\alpha\beta}$ and 4 $B_\alpha$ unknown coefficients. Proceeding with the substitution yields the following system of 10 equations: 
\\[-12pt]
\begin{subequations} \label{A-system-1}
    \begin{align}
        \label{A-system-1-gamma-1}
        \big( \gamma_1 \big)
        &\qquad
            \begin{aligned}
                &
                - \frac{\kappa}{24} \frac{\pi_a^2}{a} 
                - \frac{a}{\kappa} k^2 
                - \frac{3}{4} \frac{\pphiN{2}}{a^3}
                - \frac{3}{4} \frac{\ppsiN{2}}{a^3} 
                + \frac{3}{2} a^3 (U + V) 
                \\[4pt]
                & 
                \qquad \qquad \qquad \quad
                - \frac{\kappa}{3} \frac{\pi_a}{a^2} A_{11}
                + \frac{\pphi}{a^3} A_{13}
                + \frac{\ppsi}{a^3} A_{14} = 0 \ , 
            \end{aligned}
        \\[10pt]
        \label{A-system-1-gamma-2}
        \big( \gamma_2 \big)
        &\qquad
            \frac{a}{3\kappa} k^2 
            - \frac{\kappa}{3} \frac{\pi_a}{a^2} A_{12}
            + \frac{\pphi}{a^3} A_{23}
            + \frac{\ppsi}{a^3} A_{24} = 0 \ ,
        \\[10pt]
        \label{A-system-1-gamma-3}
        \big( \gamma_3 \big) 
        &\qquad
            a^3 \frac{\partial U}{\partial \phi} 
            - \frac{\kappa}{3} \frac{\pi_a}{a^2} A_{13} 
            + \frac{\pphi}{a^3} A_{33}
            + \frac{\ppsi}{a^3} A_{34} = 0 \ ,
        \\[10pt]
        \label{A-system-1-gamma-4}
        \big( \gamma_4 \big) 
        &\qquad
            a^3 \frac{\partial V}{\partial \psi} 
            - \frac{\kappa}{3} \frac{\pi_a}{a^2} A_{14} 
            + \frac{\pphi}{a^3} A_{34} 
            + \frac{\ppsi}{a^3} A_{44} = 0 \ ,
        \\[10pt]
        \label{A-system-1-constants}
        \big( \text{\footnotesize constants} \big) 
        &\qquad 
            - \frac{\kappa}{3} \frac{\pi_a}{a^2} B_1 
            + \frac{\pphi}{a^3} B_3 
            + \frac{\ppsi}{a^3} B_4 = 0
    \end{align}
\end{subequations}
from the vanishing scalar constraint and
\begin{subequations} \label{A-system-2}
    \begin{align}
        \label{A-system-2-gamma-1}
        \big( \gamma_1 \big) 
        &\qquad
            \frac{1}{6} a \pi_a 
            - \frac{2}{3} A_{11} 
            - 2 A_{12} = 0 \ ,
        \\[4pt]
        \label{A-system-2-gamma-2}
        \big( \gamma_2 \big) 
        &\qquad
            - \frac{2}{9} a \pi_a 
            - \frac{2}{3} A_{12} 
            - 2 A_{22} = 0 \ ,
        \\[4pt]
        \label{A-system-2-gamma-3}
        \big( \gamma_3 \big)
        &\qquad
            \pphi 
            - \frac{2}{3} A_{13} 
            - 2 A_{23} = 0 \ ,
        \\[4pt]
        \label{A-system-2-gamma-4}
        \big( \gamma_4 \big)
        &\qquad
            \ppsi 
            - \frac{2}{3} A_{14} 
            - 2 A_{24} = 0 \ ,
        \\[4pt]
        \label{A-system-2-constants}
        \big( \text{\footnotesize constants} \big)
        &\qquad
            - \frac{2}{3} B_1 
            - 2 B_2 = 0
    \end{align}
\end{subequations}
from the vanishing vector constraint.

Only 2 equations – the vanishing of the constant terms – concern the 4 $B_\alpha$ coefficients.
Therefore, the solutions for the $B_\alpha$ form a 2-dimensional space and can be written in terms of two free parameters as
\begin{equation}
    \begin{aligned}
        B_1 &= \frac{3}{\kappa} \frac{1}{a \pi_a} \big( p_\phi P_\phi + p_\psi P_\psi \big) \ , 
        \\[4pt]
        B_2 &= - \frac{1}{\kappa} \frac{1}{a \pi_a} \big( p_\phi P_\phi + p_\psi P_\psi \big) \ , 
        \\[4pt]
        B_3 &= P_\phi \ , 
        \\[4pt]
        B_4 &= P_\psi \ ,
    \end{aligned}
\end{equation}
where the dependence of the $B_\alpha$ on the free parameters $P_\phi$ and $P_\psi$ is chosen for later convenience.
Substituting the $B_\alpha$ we have found so far back into \eqref{S_P-ansatz} gives
\begin{equation} \label{S_P-with-B-alpha}
    S_{\scriptscriptstyle P} = 
    \frac{1}{2} A_{\alpha \beta} \gamma_\alpha \gamma_\beta
    + \frac{1}{\kappa \, a \pi_a} \left( \pphi P_\phi + \ppsi P_\psi \right) \left( 3 \gamma_1 - \gamma_2 \right)
    + P_\phi \, \delta \phi
    + P_\psi \, \delta \psi \ .
\end{equation}

The system of equations for $A_{\alpha\beta}$ is decoupled from that for $B_\alpha$. Therefore, $A_{\alpha\beta}$ will not have a dependence on the free parameters $P_\phi$ and $P_\psi$ we have introduced. Being so, we can already find expressions for their conjugate variables
$Q^\phi$ and $Q^\psi$ via
$Q^F = \dfrac{\partial S_{\scriptscriptstyle P}}{\partial P_F}$ where $F \in \{\phi,\psi\}$. These are
\begin{equation}
    \begin{aligned}
        Q^\phi &= \delta \phi + \frac{3 \pphi}{\kappa a \pi_a} \left( \gamma_1 - \frac{1}{3} \gamma_2 \right) \ , 
        \\[6pt]
        Q^\psi &= \delta \psi + \frac{3 \ppsi}{\kappa a \pi_a} \left( \gamma_1 - \frac{1}{3} \gamma_2 \right) \ .
        \\[4pt]
    \end{aligned}
\end{equation}
To complete the canonical transformation we need to find all the remaining relations between old and new variables by means of \eqref{pi-alpha-action}. These will express the old momenta as linear combinations of $\gamma_1, \gamma_2, \delta\phi, \delta\psi, P_\phi \text{ and } P_\psi$. Having these, all the old variables can then be rewritten as functions of the four gauge-invariant new coordinates $Q^\phi, P_\phi, Q^\psi \text{ and } P_\psi$, considering that $\gamma_1$ and $\gamma_2$ are the gauge parameters $\beta_\mu$ mentioned in Section \ref{sec:GI-DOFs} that represent gauge functions. Altogether, the old variables are given in terms of the new ones by
\begin{equation} \label{old-to-new-variables}
    \begin{aligned}
        \delta \phi &= Q^\phi + [\gamma_1, \gamma_2] \ ,
        \\[4pt]
        \delta \psi &= Q^\psi + [\gamma_1, \gamma_2] \ ,
        \\[4pt]
        \pi^1 &= \frac{3}{\kappa a \pi_a} \left( \pphi P_\phi + \ppsi P_\psi \right) + A_{13} Q^\phi + A_{14} Q^\psi + [\gamma_1, \gamma_2] \ ,
        \\[4pt]
        \pi^2 &= - \frac{1}{\kappa a \pi_a} \left( \pphi P_\phi + \ppsi P_\psi \right) + A_{23} Q^\phi + A_{24} Q^\psi + [\gamma_1, \gamma_2] \ ,
        \\[4pt]
        \delta \pphi &= P_\phi + A_{33} Q^\phi + A_{34} Q^\psi + [\gamma_1, \gamma_2] \ ,
        \\[4pt]
        \delta \ppsi &= P_\psi + A_{34} Q^\phi + A_{44} Q^\psi + [\gamma_1, \gamma_2] \ ,
        \\[4pt]
    \end{aligned}
\end{equation}
where the terms $[\gamma_1,\gamma_2]$ represent all the terms with $\gamma_1$ or $\gamma_2$. We do not need to write explicitly all these terms because they are \textit{pure gauge} and do not contribute to the \textit{physical} dynamics in the end.

The next step is to write the scalar gauge-invariant Hamiltonian $H_{\text{GI}}^s$ for the new variables $Q^F$ and $P_F$, given by the expression
\begin{equation} \label{Hamiltonian-GI-S_P}
    H_{\text{GI}}^s = H^s + \left\{ S_{\scriptscriptstyle P}, H^{(0)} \right\}_{\text{background}} \ ,
\end{equation}
which follows from the scalar part of the Hamiltonian for the perturbations $H^s$ written in terms of the new variables, and the Poisson bracket applies only to the background variables (which play the role of the external variables $\lambda$ in the analogous equation \eqref{Hamiltonian-new-Poisson-bracket}).
To start with, the second order in the expansion of the scalar constraint in \eqref{Hamiltonian-density} is given by
\begin{equation}
    \begin{split}
        \Ham^{(2)} = \,
        \frac{2 \kappa}{a^3} \left( \hod{ik} \hod{jl} - \frac{1}{2} \hod{ij} \hod{kl} \right) \delta \pi^{i j} \delta \pi^{k l}
        + \frac{1}{2 a^3} \delta \pphiN{2} + \frac{a^3}{2} \frac{\partial^2 U}{\partial \phi^2} \delta \phi^2 + \frac{a^3}{2} \hou{ij} \, \partial_i \delta \phi \, \partial_j \delta \phi
        \\[4pt]
        + \frac{1}{2 a^3} \delta \ppsiN{2} + \frac{a^3}{2} \frac{\partial^2 V}{\partial \psi^2} \delta \psi^2 + \frac{a^3}{2} \hou{ij} \, \partial_i \delta \psi \, \partial_j \delta \psi
        + \big[\delh{ij}\big] \ ,
    \end{split}
\end{equation}
where $\big[\delh{i j}\big]$ represents all the terms that involve $\delh{i j}$. To simplify calculations going further, we set $N=1$ and $N^i=0$ hereon. After a Fourier transformation and the use of \eqref{Ham-second-order-ito-densities}, \eqref{A-basis}, \eqref{delta-h-expansion} \eqref{delta-pi-expansion}, \eqref{S-gamma} and \eqref{Hamiltonian-new-Poisson-bracket}
we obtain
\begin{equation} \label{Hamiltonian-S-old-variables}
    \begin{split}
        H^s = \int \mathrm{d}^3 k \Bigg\{
        \frac{2 \kappa}{a^3} \left(-\frac{1}{6} \pi_1^2  + \frac{3}{2} \pi_2^2 \right) 
        &+ \frac{1}{2 a^3} \delta \pphiN{2}
        + \frac{a^3}{2} \left( \frac{\partial^2 U}{\partial \phi^2} + \frac{k^2}{a^2} \right) \delta \phi^2
        \\
        &+ \frac{1}{2 a^3} \delta \ppsiN{2}
        + \frac{a^3}{2} \left( \frac{\partial^2 V}{\partial \psi^2} + \frac{k^2}{a^2} \right) \delta \psi^2
        + \big[ \gamma_1, \gamma_2 \big] \Bigg\} \ ,
    \end{split}
\end{equation}
recalling that $k^2:=k_i k_j \, \delta^{ij} = k_i k_j \, a^2 \hou{ij}$ denotes the comoving wavenumber.
Note that in order to find $H^s$ we also added $\left\{ S_\gamma, H^{(0)} \right\}_\text{background}$ as prescribed by \eqref{Hamiltonian-new-Poisson-bracket}, but all the terms produced by this Poisson bracket involve $\gamma_1$ or $\gamma_2$, and are collected within $\big[ \gamma_1, \gamma_2 \big]$.
The reason why we may avoid writing the terms involving $\gamma_1$ and $\gamma_2$ explicitly is because they actually represent arbitrary gauge parameters as discussed in Section \ref{sec:GI-DOFs} (see \cite{goldberg_hamiltonian_1991} for further details) and do not appear in the final gauge-invariant Hamiltonian – they cancel out in the end%
\footnote{One can explicitly write out the terms with $\gamma_1$ and $\gamma_2$ collected in $\big[ \gamma_1, \gamma_2 \big]$ and track them throughout the complete calculation to find that in the end they cancel out of the final gauge-invariant Hamiltonian, as pointed out in \cite{Langlois:1994ec}.}%
. Thus, to simplify (greatly) the calculation for the scalar gauge-invariant sector (in fact these terms, coming from the 3-curvature, are the hardest to compute explicitly), we can simply package them into that bracket that can be ignored until it cancels out of the final expression.
Next, using \eqref{old-to-new-variables} we can rewrite \eqref{Hamiltonian-S-old-variables} in terms of $Q^F$ and $P_F$ and the yet undetermined $A_{\alpha \beta}$ as
\begin{equation} \label{Hamiltonian-S-in-A_ab}
    \begin{split}
        H^s = \int \mathrm{d}^3k 
        \frac{1}{a^3} \Bigg\{ 
        &\frac{1}{2} \left( P_\phi^2 + P_\psi^2 \right)
        \\
        &+ \left[ - \frac{2 \pphi}{a \pi_a} \left( A_{13} + 3A_{23} \right) + A_{33} \right] P_\phi Q^\phi
        + \left[ - \frac{2 \ppsi}{a \pi_a} \left( A_{13} + 3A_{23} \right) + A_{34} \right] P_\psi Q^\phi
        \\
        &+ \left[ - \frac{2 \pphi}{a \pi_a} \left( A_{14} + 3A_{24} \right) + A_{34} \right] P_\phi Q^\psi
        + \left[ - \frac{2 \ppsi}{a \pi_a} \left( A_{14} + 3A_{24} \right) + A_{44} \right] P_\psi Q^\psi
        \\
        &+ \left[ - \frac{2 \kappa}{3} A_{13} A_{14} + 6 \kappa A_{23} A_{24} + A_{33} A_{34} + A_{34} A_{44} \right] Q^\phi Q^\psi
        \\
        &+ \left[ - \frac{\kappa}{3} A_{13}^2 + 3 \kappa A_{23}^2 + \frac{1}{2} A_{33}^2 + \frac{1}{2} A_{34}^2 + \frac{1}{2} a^6 \left( \frac{\partial^2 U}{\partial \phi^2} + \frac{k^2}{a^2} \right) \right] \left(Q^\phi\right)^2
        \\
        &+ \left[ - \frac{\kappa}{3} A_{14}^2 + 3 \kappa A_{24}^2 + \frac{1}{2} A_{34}^2 + \frac{1}{2} A_{44}^2 + \frac{1}{2} a^6 \left( \frac{\partial^2 V}{\partial \psi^2} + \frac{k^2}{a^2} \right) \right] \left(Q^\psi\right)^2
        \\
        &+ \big[ \gamma_1, \gamma_2 \big]
        \Bigg\} \ .
        \\[-20pt]
    \end{split}
\end{equation}

At this point, let us turn our attention to the seven coefficients $A_{13}$, $A_{14}$, $A_{23}$, $A_{24}$, $A_{33}$, $A_{34}$ and $A_{44}$ that appear explicitly in \eqref{Hamiltonian-S-in-A_ab}.
We obtained 8 equations for the $A_{\alpha \beta}$ coefficients by recognising that the coefficients for all perturbation variables in $E$ and $M$ must vanish. It turns out that one of these equations is a linear combination of the others and the vanishing background scalar constraint \eqref{Background-scalar-constraint}
\footnote{
Precisely, $\dfrac{2}{3} \eqref{A-system-1-gamma-1} + 2 \eqref{A-system-1-gamma-2} + \dfrac{\pphi}{a^3} \eqref{A-system-2-gamma-3} + \dfrac{\ppsi}{a^3} \eqref{A-system-2-gamma-4} - \dfrac{\kappa \pi_a}{3a^2} \eqref{A-system-2-gamma-1} = \Hamorder{0} = 0$.
}.
Thus, there are only 7 independent equations constraining the 10 different $A_{\alpha\beta}$ coefficients. We therefore have 3 free conditions which we can arbitrarily impose on the $A_{\alpha\beta}$ and still recover an appropriate generating function. This can be understood as the freedom to redefine
\begin{equation} \label{P_F-redefinition}
    P_\phi \longrightarrow P_\phi +f_1(t) \, Q^\phi +f_2(t) \, Q^\psi
    \qquad \text{ and } \qquad
    P_\psi \longrightarrow P_\psi +f_3(t) \, Q^\phi +f_4(t) \, Q^\psi \ ,
\end{equation}
yet with the different $f_i$ constrained by $\big\{ P_\phi , P_\psi \big\} = 0$ to preserve the Poisson structure.
We shall exploit this freedom by imposing 3 extra conditions that will simplify $H^s$ conveniently, given by
\begin{equation} \label{A-expressions-1}
    A_{33} = \frac{3 \pphiN{2}}{a \pi_a} \ ,
    \qquad \qquad
    A_{34} = \frac{3 \pphi \ppsi}{a \pi_a} \ ,
    \qquad \text{and} \qquad
    A_{44} = \frac{3 \ppsiN{2}}{a \pi_a} \ .
\end{equation}
As a result, the coefficients of the $P_F Q^{F'}$ cross-terms in \eqref{Hamiltonian-S-in-A_ab} vanish in light of equations ($\gamma_3$) and ($\gamma_4$) in \eqref{A-system-2}.
Fixing these extra conditions renders the system for the $A_{\alpha \beta}$ fully determinable. From equations ($\gamma_3$) and ($\gamma_4$) in \eqref{A-system-1}, and ($\gamma_3$) and ($\gamma_4$) in \eqref{A-system-2} we obtain the remaining coefficients that concern us, that is
\begin{equation} \label{A-expressions-2}
    \begin{aligned}
        A_{13} &= \frac{3 a^5}{\kappa \pi_a} \frac{\partial U}{\partial \phi} + \frac{9\, \pphiN{3}}{\kappa\, a^2 \pi_a^2} + \frac{9\, \pphi \ppsiN{2}}{\kappa\, a^2 \pi_a^2} \ ,
        \\[2pt]
        A_{14} &= \frac{3 a^5}{\kappa \pi_a} \frac{\partial V}{\partial \psi} + \frac{9\, \ppsi^3}{\kappa\, a^2 \pi_a^2} + \frac{9\, \ppsi \pphiN{2}}{\kappa\, a^2 \pi_a^2} \ ,
        \\[3pt]
        A_{23} &= \frac{1}{2} \pphi - \frac{a^5}{\kappa \pi_a} \frac{\partial U}{\partial \phi} - \frac{3}{\kappa} \frac{\pphi}{a^2 \pi_a^2} \left(\pphiN{2} + \ppsiN{2} \right) \ ,
        \\[2pt]
        A_{24} &= \frac{1}{2} \ppsi - \frac{a^5}{\kappa \pi_a} \frac{\partial V}{\partial \psi} - \frac{3}{\kappa} \frac{\ppsi}{a^2 \pi_a^2} \left(\pphiN{2} + \ppsiN{2} \right) \ .
    \end{aligned}
\end{equation}
We can now substitute all the coefficients in \eqref{A-expressions-1} and \eqref{A-expressions-2} back into \eqref{Hamiltonian-S-in-A_ab} and find
\begin{equation} \label{H-S-final-expression}
    \begin{split}
        H^s = \int d^3k \, \frac{1}{2a^3} 
        \Bigg\{ 
            & P_\phi^2 + P_\psi^2 
            + \left[ 
                3 \kappa \, \pphi \ppsi 
                - 6 \frac{a^5}{\pi_a} \left( \pphi \frac{\partial V}{\partial \psi} + \ppsi \frac{\partial U}{\partial \phi} \right) 
            \right] Q^\phi Q^\psi
            \\
            & + \left[ 
                \frac{3\kappa}{2} \pphiN{2} 
                - 6 \frac{a^5 \pphi}{\pi_a} \frac{\partial U}{\partial \phi} 
                - 9 \frac{\pphiN{2}}{a^2 \pi_a^2} \left( \pphiN{2} + \ppsiN{2} \right) 
                + a^6 \left( \frac{\partial^2 U}{\partial \phi^2} + \frac{k^2}{a^2} \right) 
            \right] Q^{\phi \, 2}
            \\
            & + \left[ 
                \frac{3\kappa}{2} \ppsiN{2} 
                - 6 \frac{a^5 \ppsi}{\pi_a} \frac{\partial V}{\partial \psi} 
                - 9 \frac{\ppsiN{2}}{a^2 \pi_a^2} \left( \pphiN{2} + \ppsiN{2} \right) 
                + a^6 \left( \frac{\partial^2 V}{\partial \psi^2} + \frac{k^2}{a^2} \right) 
            \right] Q^{\psi \, 2}
            \\
            & + \big[ \gamma_1, \gamma_2 \big]
        \Bigg\} \ .
        \\[-20pt]
    \end{split}
\end{equation}

To find the gauge-invariant scalar Hamiltonian prescribed by \eqref{Hamiltonian-GI-S_P}, we only have the Poisson bracket left to compute. 
It will follow from the generating function \eqref{S_P-ansatz} with all the $A_{\alpha \beta}$ and $B_\alpha$ that we have now explicitly computed, and the background Hamiltonian \eqref{Hamiltonian-0-notdensity} recalling we had set $N=1$ earlier. This Poisson bracket is found to be
\begin{equation} \label{Poisson-bracket-for-H-GI}
    \left\{ S_{\scriptscriptstyle P} , H^{(0)} \right\}_\text{background}
    = \frac{1}{2} \dot{A}_{33} \, \delta \phi^2
    + \dot{A}_{34} \, \delta \phi \delta \psi
    + \frac{1}{2} \dot{A}_{44} \, \delta \psi^2
    + \left[ \gamma_1, \gamma_2 \right] \ .
\end{equation}
where $\dot{A}_{\alpha\beta} \equiv \big\{ A_{\alpha\beta} , H^{(0)} \big\}_\text{background}$ .
Finally, we can substitute \eqref{H-S-final-expression} and \eqref{Poisson-bracket-for-H-GI} into \eqref{Hamiltonian-GI-S_P} and obtain the gauge-invariant scalar Hamiltonian
\begin{equation} \label{H-S-GI-final}
    \begin{split}
        H^s_{\text{GI}} = \int d^3k \, \frac{1}{a^3} \Bigg\{ 
        &\frac{1}{2} P_\phi^2 + \frac{1}{2} P_\psi^2 
        + \left[ 3 \kappa \, \pphi \ppsi - 9 \frac{\pphi \ppsi}{a^2 \pi_a^2} \left(\pphiN{2} + \ppsiN{2} \right) - 6 \frac{a^5}{\pi_a} \left(\ppsi \frac{\partial U}{\partial \phi} + \pphi \frac{\partial V}{\partial \psi}\right) \right] Q^\phi Q^\psi
        \\
        &+ \frac{1}{2} \left[ 3\kappa \, \pphiN{2} - 18 \frac{\pphiN{2}}{a^2 \pi_a^2} \left(\pphiN{2} + \ppsiN{2} \right) - 12 \frac{a^5 \pphi}{\pi_a} \frac{\partial U}{\partial \phi} + a^6 \left(\frac{\partial^2 U}{\partial \phi^2} + \frac{k^2}{a^2} \right) \right] Q^{\phi \, 2}
        \\
        &+ \frac{1}{2} \left[ 3\kappa \, \ppsiN{2} - 18 \frac{\ppsiN{2}}{a^2 \pi_a^2} \left(\pphiN{2} + \ppsiN{2} \right) - 12 \frac{a^5 \ppsi}{\pi_a} \frac{\partial V}{\partial \psi} + a^6 \left(\frac{\partial^2 V}{\partial \psi^2} + \frac{k^2}{a^2} \right) \right] Q^{\psi \, 2}
        \Bigg\} \ .
        \\[4pt]
    \end{split}
\end{equation}
As was pointed out earlier, all the terms involving $\gamma_1$ and $\gamma_2$ cancelled out with this last substitution, justifying the choice to ignore these terms altogether. This does not come as a surprise if one considers that arbitrary gauge functions should not appear in the resulting gauge-invariant Hamiltonian.

To conclude, from Hamilton's equations
\begin{equation} \label{Hamilton-equations-scalar}
    \dot{Q}^F = \frac{\delta H_{GI}^s}{\delta P_F}
    \qquad \text{and} \qquad
    \dot{P}_F = -\frac{\delta H_{GI}^s}{\delta Q^F} \ ,
\end{equation}
we can obtain the equations of motion for the scalar gauge-invariant variables, which are found to be
\begin{equation} \label{Q_phi_EOM}
    \begin{split}
        \ddot{Q}^\phi + 3 \frac{\dot{a}}{a} \dot{Q}^\phi + \Bigg[ 
        & 3 \kappa \frac{\pphi \ppsi}{a^6} - 9 \frac{\pphi \ppsi}{a^8 \pi_a^2} \left(\pphiN{2} + \ppsiN{2}\right) - 6 \frac{1}{a \pi_a} \left(\ppsi \frac{\partial U}{\partial \phi} + \pphi \frac{\partial V}{\partial \psi}\right) \Bigg] Q^\psi
        \\[6pt]
        + \Bigg[ 
        & 3 \kappa \frac{\pphiN{2}}{a^6} - 18 \frac{\pphiN{2}}{a^8 \pi_a^2} \left(\pphiN{2} + \ppsiN{2}\right) - 12 \frac{\pphi}{a \pi_a} \frac{\partial U}{\partial \phi} + \frac{\partial^2 U}{\partial \phi^2} + \frac{k^2}{a^2} \Bigg] Q^{\phi} = 0 \ ,
    \end{split}
\end{equation}
and
\begin{equation} \label{Q_psi_EOM}
    \begin{split}
        \ddot{Q}^\psi + 3 \frac{\dot{a}}{a} \dot{Q}^\psi + \Bigg[ 
        & 3 \kappa \frac{\pphi \ppsi}{a^6} - 9 \frac{\pphi \ppsi}{a^8 \pi_a^2} \left(\pphiN{2} + \ppsiN{2}\right) - 6 \frac{1}{a \pi_a} \left(\ppsi \frac{\partial U}{\partial \phi} + \pphi \frac{\partial V}{\partial \psi}\right) \Bigg] Q^\phi
        \\[6pt]
        + \Bigg[ 
        & 3 \kappa \frac{\ppsiN{2}}{a^6} - 18 \frac{\ppsiN{2}}{a^8 \pi_a^2} \left(\pphiN{2} + \ppsiN{2}\right) - 12 \frac{\ppsi}{a \pi_a} \frac{\partial V}{\partial \psi} + \frac{\partial^2 V}{\partial \psi^2} + \frac{k^2}{a^2} \Bigg] Q^{\psi} = 0 \ .
    \end{split}
\end{equation}
These can be collectively recasted in a succinct way like
\begin{equation} \label{QF-EOM}
    \ddot{Q}^F + 3 \frac{\dot{a}}{a} \dot{Q}^F + \frac{k^2}{a^2} Q^F + \sum_{F'} \frac{\Omega^2_{F F'}}{a^2} Q^{F'} = 0 \ ,
\end{equation}
where $F$ and $F'$ run over the space of scalar fields present ($\{\phi,\psi\}$ so far), and $\Omega^2_{F F'}$ is given by
\begin{equation} \label{Omega_FF'}
    \Omega^2_{FF'} = 
    3\kappa \frac{p_{_{\!F}} p_{_{\!F'}}}{a^4} 
    - 9 \left( 1 + \delta_{_{\!FF'}} \right) \frac{p_{_{\!F}} p_{_{\!F'}}}{a^6 \pi_a^2} \sum_{_{\!F''}} p_{_{\!F''}}^2 
    - 6 \frac{a}{\pi_a} \left( p_{_{\!F}} \frac{\partial U_{_{\!F'}}}{\partial F'} + p_{_{\!F'}} \frac{\partial U_{_{\!F}}}{\partial F} \right) 
    + \delta_{_{\!FF'}} a^2 \, \frac{\partial^2 U_{_{\!F}}}{\partial F^2} \ ,
\end{equation}
for which $U_{_F}$ is short-hand for the potential associated to the field $F$.
The equations of motion of the gauge-invariant variables $Q^F$ are coupled via the terms with $\Omega^2_{FF'}$. Note that this factor depends only on background quantities.
Their coupled evolution may lead to new phenomenology that cannot be inferred by simply superimposing the behaviour of two perturbation variables evolved neglecting the coupling that arises through their interaction with the gravitational field.
An example of such new phenomenology can be found in \cite{Frion:2025}. In the following section we will further generalise the formalism prescribed so far to the case with multiple scalar fields without much complication. This is the main result of this work.

\section{Generalisation to multiple scalar fields}
We now wish to extend the previous analysis to a cosmological setting with $n_f$ scalar fields minimally coupled to gravity.
Up to linear order, this extension is not complicated and simply amounts to linearly-extending the relevant expressions for $n_f$ fields.
However, more care is required once the second-order Hamiltonian enters the picture.
We will summarise the key points involved in the extension in terms of the different degrees of freedom, the system of equations for the coefficients of the generating function, and the extra conditions available to simplify the system conveniently.

In this setting, the scalar subspace of linear perturbations is comprised of $(n_f+2)$ coordinates: $n_f$ linear perturbations of the scalar fields and 2 scalar components of the linear perturbations of the spatial metric. We denote these collectively as $\big\{ \gamma_\alpha \big\}$ for $\alpha=1,...,n_f+2$.
Hence, the phase space of scalar perturbations is spanned by $ 2(n_f + 2) $ degrees of freedom.

As before, to extract the gauge-invariant sector from the scalar phase space
we consider a canonical transformation generated by the quadratic function
\begin{equation} \label{gen_function}
    S_\gamma (\gamma_\alpha,P_\alpha) = \frac{1}{2} A_{\alpha\beta}\, \gamma_\alpha \gamma_\beta + B_\alpha\, \gamma_\alpha \ ,
\end{equation}
where recall that the coefficient matrix $A_{\alpha\beta}$ is symmetric and integration over $\mathbf{k}$ is implied.
The number of unknown coefficients that we need to determine to perform the canonical transformation are 
$\frac{1}{2}\,(n_f+2)(n_f+3)$ for $A_{\alpha\beta}$ coefficients
and
$(n_f+2)$ for $B_\alpha$ coefficients.
The linearised scalar and vector constraints in this setting (i.e., analogous to $E$ and $M$)
are polynomials of linear order in the perturbation variables $\gamma_\alpha$, each with $(n_f+2)$ coefficients for the $\gamma_\alpha$ and 1 constant term.
In order for the constraints to be satisfied, each of the $(n_f + 3)$ coefficients must vanish identically.
Hence, in total, the constraints supply $2(n_f + 3)$ equations that must be satisfied by the $A_{\alpha\beta}$ and $B_\alpha$ coefficients so that transformation incurred is indeed canonical and preserves the Poisson structure. 

Similar to the two-field case, the system of equations for the $A_{\alpha\beta}$ and $B_\alpha$ is decoupled, such that the $B_\alpha$ can be determined separately from the $A_{\alpha\beta}$.
Regarding the $B_\alpha$ system,
we only have the 2 equations given by the vanishing of the constant to determine $(n_f+2)$ unknowns. 
Consequently, the general solution for the $B_\alpha$ coefficients is an $n_f$-dimensional space which we shall parametrise by introducing $n_f$ free parameters, denoted by $P_F$ for $F=1,...,n_f$. 
In turn, these result in $n_f$ gauge-invariant variables via the relation
\begin{equation} \label{gauge_invariant_Q}
    Q^F = \frac{\partial S_\gamma}{\partial P_F} \ ,
\end{equation}
with each $Q^F$ associated to one linear perturbation of a scalar field $\delta F$.

Regarding the $A_{\alpha\beta}$ system, 
the remaining $(2n_f+4)$ equations are not independent as one of them can be reexpressed as a linear combination of the others and the energy constraint. Only $(2n_f+3)$ equations are linearly independent.
These equations constrain the $\frac{1}{2}(n_f+2)(n_f+3)$ coefficients in $A_{\alpha\beta}$, such that the difference, that is $\frac{1}{2}n_f(n_f+1)$, represents the number of free conditions that can be imposed on the $A_{\alpha\beta}$ without affecting the Poisson structure.
This freedom can be understood as the possibility to redefine the momenta $P_F$ (analogous to \eqref{P_F-redefinition}) via
\begin{equation}
    P_F \longrightarrow P_F + \sum_{F'=1}^{n_f} f_{FF'}(t) \, Q^{F'} \ ,
\end{equation}
where the functions $f_{FF'}(t)$, with $F,F'=1,\dots,n_f$, are arbitrary up to the requirement that the Poisson brackets are preserved, namely 
\begin{equation} \label{poisson_preservation}
    \big\{ P_F, P_{F'} \big\} = 0 \ .
\end{equation}
Since this condition imposes $\frac{1}{2}\,n_f\,(n_f-1)$ constraints on the $n_f^2$ functions $f_{FF'}(t)$, the net number of free functions available is 
$n_f^2 - \frac{1}{2}\,n_f\,(n_f-1) = \frac{1}{2}\,n_f\,(n_f+1)$ ,
which coincides with the number of free conditions that can be added to the $A_{\alpha\beta}$ system. 
Remarkably, the number of free conditions matches the number of degrees of freedom in the coefficients of the $n_f^2$ cross-terms of the form $P_F\, Q^{F'}$ in the scalar Hamiltonian. That means that these free conditions can be chosen appropriately to eliminate all the cross-terms from the scalar Hamiltonian.

Once these simplifying conditions are added, the system for the $A_{\alpha\beta}$ is fully determined, leading to a Hamiltonian that contains only the quadratic terms ${P_F}^2$ and ${Q^F}^2$, and interaction terms of the form $Q^F Q^{F'}$. In the process of solving for the coefficients $A_{\alpha\beta}$ with the full system of equations, one obtains expressions that generalise in a straightforward manner the two-field case.
Recalling that the values of the indices in $A_{\alpha\beta}$ represent the different scalar field's perturbations ($\alpha=3,...,n_f$) and the two scalar components of the metric perturbations ($\alpha=1,2$), a given $A_{\alpha\beta}$ will have a form analogous to one of those given in \eqref{A-expressions-1} and \eqref{A-expressions-2} after adjusting its fields, their potentials and their momenta to match the indices $\alpha$ and $\beta$. However, the not-so-straightforward generalisation is that any occurrence of the term $(p_\phi^2 + p_\psi^2)$, like in $A_{23}$ and $A_{24}$ in \eqref{A-expressions-2}, is in fact the quadrature sum of \textit{all} the scalar field's momenta (not just the fields relevant to the indices), and thus must be replaced accordingly by
$\sum_{F=1}^{n_f} p_{\scriptscriptstyle F}^{\,\,2}$ .

Having worked out the gauge-invariant scalar Hamiltonian, all that remains is to employ Hamilton's equations to obtain the equations of motion for the $n_f$ gauge-invariant variables $Q^F$. Remarkably, the resulting equations of motion are collectively given by the same expressions as in \eqref{QF-EOM} and \eqref{Omega_FF'} except for the fact that in this case the indices $F,F'=1,...,n_f$ and not only $\phi$ and $\psi$ (i.e. 1 and 2) like in the two-field case. Explicitly, the equations of motion are
\begin{equation} \label{QF-EOM-multifield}
    \ddot{Q}^F + 3 \frac{\dot{a}}{a} \dot{Q}^F + \frac{k^2}{a^2} Q^F + \sum_{F'} \frac{\Omega^2_{F F'}}{a^2} Q^{F'} = 0
\end{equation}
where $\Omega^2_{F F'}$, for which $F,F'=1,...,n_f$, is given by
\begin{equation} \label{Omega_FF'-multifield}
    \Omega^2_{FF'} = 
    3\kappa \frac{p_{_{\!F}} p_{_{\!F'}}}{a^4} 
    - 9 \left( 1 + \delta_{_{\!FF'}} \right) \frac{p_{_{\!F}} p_{_{\!F'}}}{a^6 \pi_a^2} \sum_{_{\!F''}} p_{_{\!F''}}^2 
    - 6 \frac{a}{\pi_a} \left( p_{_{\!F}} \frac{\partial U_{_{\!F'}}}{\partial F'} + p_{_{\!F'}} \frac{\partial U_{_{\!F}}}{\partial F} \right) 
    + \delta_{_{\!FF'}} a^2 \, \frac{\partial^2 U_{_{\!F}}}{\partial F^2} \; .
\end{equation}
This result is in agreement with other works in the literature such as \cite{Gong:2016qmq}.
Lastly, one can see that in the case of a single field (i.e. $F,F'=\phi$), the expression found by Langlois \cite{Langlois:1994ec} is recovered%
\footnote{When comparing this work to \cite{Langlois:1994ec} one should note that there $k^2 = k_i k_j \hou{ij}$, whereas here we defined $k^2=k_i k_j \delta^{ij} = a^2 k_i k_j \hou{ij}$.}%
.

\section{Tensor sector} \label{sec:Tensor}
Let us now turn our attention to the tensor sector. 
Tensor modes correspond to gravitational waves and can be present even if there is no matter, unlike the case for scalar modes. 
Since there is no tensor component in the linearised constraints, the conjugate coordinates $\gamma_m$ and $\pi_m$ for $m=5,6$ are already gauge-invariant right away. The evolution of the tensor sector, as we will see at the end, is not affected by the presence of multiple scalar fields. To simplify our calculations without loss of generality, we will focus again on the two-field case in this section.
We will first express the full second-order Hamiltonian. Then, we will extract its tensor sector in terms of the new coordinates by means of adequate canonical transformations. In the end we will employ Hamilton's equations to obtain the new variables' equations of motion.

In order to preserve the canonical Poisson structure, the canonical transformation of the variables must be obtained using the generating function \eqref{S-gamma} via
\begin{equation} \label{H-2_gamma-transformation}
    H_\gamma^{(2)}[\gamma_m,\pi_m] = H^{(2)} + \Big\{ S_\gamma, H^{(0)} \Big\} \ .
\end{equation}
Regarding the second term, recalling the $a^2$ dependence of $S_\gamma$ via the matrices $A_{ij}^m$ we then find
\begin{equation}
    \Big\{ S_\gamma, H^{(0)} \Big\} = \frac{\kappa \pi_a}{3 a^2} \sum_{m=1}^6 \pi_m \gamma_m \ .
\end{equation}
With regards to the first term, after setting $N=1$ and $N_i=0$ right away in \eqref{Hamiltonian} for simplicity, the second-order Hamiltonian is given by
\begin{equation}
    H^{(2)} = \int \mathrm{d}^3x \; \Hamorder{2} \ .
\end{equation}
In the tensor case, the terms coming from the 3-curvature will be relevant, unlike in the scalar case where they were pure gauge and could be ignored. We therefore express again here the second-order scalar constraint with all terms written explicitly, that is
\begin{align} \label{Hamiltonian-density-2-tensor}
        \Hamorder{2} 
        = & \; \frac{\kappa}{a^3}
        \Biggl[
            \frac{1}{48} a^2 \pi_a^2
            \left( \frac{5}{3} \hou{ik} \hou{jl} - \frac{1}{2} \hou{ij} \hou{kl} \right)
            \delh{ij} \delh{kl}
            \nonumber \\
            & \qquad
            + \frac{1}{3} a \pi_a \delh{ij} \delpi{ij}
            - \frac{1}{6} a \pi_a \hou{ij} \hod{kl} \, \delh{ij} \delpi{kl}
            + 2 \left( \hod{ik} \hod{jl} - \frac{1}{2} \hod{ij} \hod{kl} \right) \delpi{ij} \delpi{kl}
        \Biggr]
        \nonumber \\
        & - \frac{a^3}{8\kappa} 
            \Biggl[
                \delh{ij} \, \partial_m \partial_n \delh{kl}
                \left(
                    4 \hou{in} \hou{jm} \hou{kl}
                    + 4 \hou{ik} \hou{jl} \hou{mn}
                    - 8 \hou{il} \hou{jm} \hou{kn}
                    + 2 \hou{ij} \hou{km} \hou{ln}
                    - 2 \hou{ij} \hou{kl} \hou{mn}
                \right)
                \nonumber \\
                & \qquad \quad
                + \partial_m \delh{ij} \, \partial_n \delh{kl}
                \left(
                    -4 \hou{im} \hou{jl} \hou{kn}
                    + 4 \hou{im} \hou{jn} \hou{kl}
                    + 3 \hou{ik} \hou{jl} \hou{mn}
                    - \hou{ij} \hou{kl} \hou{mn}
                    - 2 \hou{ik} \hou{jn} \hou{lm}
                \right)
            \Biggr]
        \nonumber \\ 
        & + \frac{1}{a^3}
            \Biggl[
                \frac{1}{2} \delta \pphiN{2}
                - \frac{1}{2} \pphi  \hou{ij} \delh{ij} \delta \pphi
                + \frac{1}{8} \pphiN{2} \hou{il} \hou{jk} \delh{ij} \delh{kl}
                + \frac{1}{16}  \pphiN{2} \hou{ij} \hou{kl} \delh{ij} \delh{kl}
            \Biggr]
        \nonumber \\
        & + a^3 \,
            \Biggl[
                \frac{1}{2} \frac{\partial^2 U}{\partial \phi^2} \delta \phi^2
                + \frac{1}{2} \frac{\partial U}{\partial \phi} \hou{ij} \delh{ij} \delta \phi
                - \frac{1}{4} U(\phi) \Bigl(\hou{il} \hou{jk} - \frac{1}{2} \hou{ij} \hou{kl}\Bigr) \delh{ij} \delh{kl}
            \Biggr]
        \nonumber \\[4pt]
        & + \frac{1}{2} a^3 \hou{ij} \partial_i \delta \phi \, \partial_j \delta \phi
        \nonumber \\[6pt]
        & + \frac{1}{a^3}
            \Biggl[
                \frac{1}{2} \delta \ppsiN{2}
                - \frac{1}{2} \ppsi  \hou{ij} \delh{ij} \delta \ppsi
                + \frac{1}{8} \ppsiN{2} \hou{il} \hou{jk} \delh{ij} \delh{kl}
                + \frac{1}{16}  \ppsiN{2} \hou{ij} \hou{kl} \delh{ij} \delh{kl}
            \Biggr]
        \nonumber \\
        & + a^3 \,
            \Biggl[
                \frac{1}{2} \frac{\partial^2 V}{\partial \psi^2} \delta \psi^2
                + \frac{1}{2} \frac{\partial V}{\partial \psi} \hou{ij} \delh{ij} \delta \psi
                - \frac{1}{4} V(\psi) \Bigl(\hou{il} \hou{jk} - \frac{1}{2} \hou{ij} \hou{kl}\Bigr) \delh{ij} \delh{kl}
            \Biggr]
        \nonumber \\[4pt]
        & + \frac{1}{2} a^3 \hou{ij} \partial_i \delta \psi \, \partial_j \delta \psi \ .
\end{align}
After performing a Fourier transform, the second-order Hamiltonian is then found to be
\begin{equation} \label{Hamiltonian-2-k-integral}
    \begin{split}
        H^{(2)}
        = \int \mathrm{d}^3k \,
        \Biggl\{
            & \frac{\kappa}{a^3}
            \biggl[
                \frac{1}{48} a^2 \pi_a^2 \delh{ij} \delh{kl}
                \left(
                    \frac{5}{3} \hou{ik} \hou{jl}
                    - \frac{1}{2} \hou{ij} \hou{kl}
                \right)
                + \frac{1}{3} a \pi_a \delpi{ij} \delh{ij}
                \\
                & \qquad \quad
                - \frac{1}{6} a \pi_a \hou{ij} \hod{kl} \delpi{kl} \delh{ij}
                + 2 \left( \hod{ik} \hod{jl} - \frac{1}{2} \hod{ij} \hod{kl} \right)
                \delpi{ij} \delpi{kl}
            \biggr]
            \\[6pt]
            &
            + \frac{a^3}{8\kappa}
            k_m k_n
            \delh{ij} \delh{kl}
            \left[
                \hou{ik} \hou{jl} \hou{mn}
                - 2 \hou{il} \hou{jm} \hou{kn} + 2 \hou{ij} \hou{km} \hou{ln}
                - \hou{ij} \hou{kl} \hou{mn}
            \right]
            \\[6pt]
            &
            + \frac{1}{a^3}
            \left[
                \frac{1}{2} \delta \pphiN{2}
                - \frac{1}{2} \pphi \hou{ij} \delh{ij} \delta \pphi
                + \frac{1}{8} \pphiN{2} \hou{il} \hou{jk} \delh{ij} \delh{kl}
                + \frac{1}{16} \pphiN{2} \hou{ij} \hou{kl} \delh{ij} \delh{kl}
            \right]
            \\[6pt]
            &
            + a^3
            \left[
                \frac{1}{2}
                \left( \frac{\partial^2 V}{\partial \phi^2} + \frac{k^2}{a^2} \right)
                \delta \phi^2
                + \frac{1}{2} \hou{ij}
                  \frac{\partial V}{\partial \phi}
                  \delh{ij} \delta \phi
                - \frac{1}{4}
                  V(\phi)
                  \left(
                      \hou{il} \hou{jk}
                      - \frac{1}{2} \hou{ij} \hou{kl}
                  \right)
                \delh{ij} \delh{kl}
            \right]
            \\[6pt]
            &
            + \frac{1}{a^3}
            \left[
                \frac{1}{2} \delta \ppsiN{2}
                - \frac{1}{2} \ppsi \hou{ij} \delh{ij} \delta \ppsi
                + \frac{1}{8} \ppsiN{2} \hou{il} \hou{jk} \delh{ij} \delh{kl}
                + \frac{1}{16} \ppsiN{2} \hou{ij} \hou{kl} \delh{ij} \delh{kl}
            \right]
            \\[6pt]
            &
            + a^3
            \left[
                \frac{1}{2} \left( \frac{\partial^2 V}{\partial \psi^2} + \frac{k^2}{a^2} \right) \delta \psi^2
                + \frac{1}{2} \hou{ij}
                  \frac{\partial V}{\partial \psi}
                  \delh{ij} \delta \psi
                - \frac{1}{4} V(\psi) \left( \hou{il} \hou{jk} - \frac{1}{2} \hou{ij} \hou{kl} \right)
                \delh{ij} \delh{kl}
            \right]
        \Biggr\}
    \end{split}
\end{equation}
where the dependence on $\mathbf{k}$ and the tilde above of all the Fourier transformed perturbations have been omitted to prevent further cluttering of the notation.
To extract the tensor sector from \eqref{Hamiltonian-2-k-integral} we first need to rewrite it in terms of $\gamma_m$ and $\pi_m$ and recall that $m=5,6$ denote the tensor components. To do this we make use of \eqref{A-basis} – \eqref{Pi-scalar} and find
\begin{equation} \label{Hamiltonian-2-tensor-gamma-pi}
    \begin{split}
        H_{t}^{(2)} =
        \int d^3 k \sum_{m=5,6} \Biggl\{ 
            \frac{2 \kappa}{a^3} \pi_m^2
            + \frac{\kappa \pi_a}{3 a^2} \pi_m \gamma_m
            + \frac{1}{4} \biggl[ 
                & \frac{5 \kappa \pi_a^2}{36 a}
                + \frac{\pphiN{2}}{2 a^3}
                - a^3 U(\phi)
            \biggr.
        \\
        & \, \biggl.
            + \frac{\ppsiN{2}}{2 a^3}
            - a^3 V(\psi)
            + \frac{a}{2\kappa} k^2
        \biggr] 
        \gamma_m^2
        \Biggr\} \ .
    \end{split}
\end{equation}
We then find the complete tensor part of the second-order Hamiltonian given by \eqref{H-2_gamma-transformation} to be
\begin{equation} \label{Hamiltonian-2-tensor-GI}
    \begin{split}
        H^t_{GI} =
        \int d^3 k \sum_{m=5,6} \Biggl\{ 
            \frac{2 \kappa}{a^3} \pi_m^2
            + \frac{2 \kappa \pi_a}{3 a^2} \pi_m \gamma_m
            + \frac{1}{4} \biggl[ 
                & \frac{5 \kappa \pi_a^2}{36 a}
                + \frac{\pphiN{2}}{2 a^3}
                - a^3 U(\phi)
            \biggr.
        \\
        & \, \biggl.
            + \frac{\ppsiN{2}}{2 a^3}
            - a^3 V(\psi)
            + \frac{a}{2\kappa} k^2
        \biggr] 
        \gamma_m^2
        \Biggr\} \ .
    \end{split}
\end{equation}

Furthermore, the Hamiltonian can be diagonalised with another canonical transformation provided by the generating function
\begin{equation} \label{tensor-final-transformation}
    S_\mathcal{T} ( \gamma_m, P_\mathcal{T}^m ) =  \sum_{m=5,6}-\frac{1}{12} a \pi_a \gamma_m^2 + P_\mathcal{T}^m \gamma_m
\end{equation}
with integration over $\mathbf{k}$ implied.
This transformation introduces new momenta variables $P_\mathcal{T}^m$ (for $m=5,6$), for which their conjugate variables are found to be
\begin{equation}
    \mathcal{T}_m = \frac{\partial S_\mathcal{T}}{\partial P_\mathcal{T}^m} = \gamma_m \ .
\end{equation}
Hence, in terms of the original variables, the new momenta $P_\mathcal{T}^m$ are
\begin{equation}
    P_\mathcal{T}^m 
    = \pi_m + \frac{1}{6} a \pi_a \gamma_m \ .
\end{equation}
This last canonical transformation then yields the simpler, diagonal form
\begin{equation}
    \tilde{H}^{t}_{GI} =
        \int d^3 k \sum_{m=5,6} \biggl\{
            \frac{2 \kappa}{a^3} (P_\mathcal{T}^m)^2
            + \frac{a}{8 \kappa} k^2 (\mathcal{T}_m  )^2
        \biggr\} \ .
\end{equation}

Finally, by means of Hamilton's equations
\begin{equation} \label{Hamilton-equations-tensor}
    \dot{\mathcal{T}}_m = \frac{\partial \tilde{H}_{GI}^t}{\partial P_\mathcal{T}^m}
    \qquad \text{and} \qquad
    \dot{P}_\mathcal{T}^m = -\frac{\partial \tilde{H}_{GI}^t}{\partial \mathcal{T}_m} \, ,
\end{equation}
we obtain the equation of motion for the two tensor modes
\begin{equation}
    \ddot{\mathcal{T}}_m + 3 \frac{\dot{a}}{a} \dot{\mathcal{T}}_m + \frac{k^2}{a^2} \mathcal{T}_m = 0 \qquad \text{ for } \ m=5,6 \ .
\end{equation}
This equation of motion resembles that of a massless scalar mode as seen by comparing it to \eqref{QF-EOM} where setting the field's potential to zero (i.e. massless) results in $\Omega^2_{F F'} = 0$.
In this sense, the two tensor modes resemble two massless scalar fields.
We close this section by pointing out that due to the decoupling of the different sectors, the result for the two tensor modes is independent of the number of scalar fields (and vector fields for that matter) minimally coupled to gravity in the scenario being considered. Hence, the result for the evolution of tensor modes trivially generalises to the case with multiple scalar fields present.
The situation for tensor modes would be different, however, if matter of `tensor' type were coupled to gravity.

\section{Conclusion}
In summary, by building upon the seminal work by Langlois \cite{Langlois:1994ec}, 
in this paper we have prescribed a Hamiltonian formalism to obtain the gauge-invariant scalar and tensor perturbations and their equations of motion for a system with multiple scalar fields comprising its matter sector.
We explicitly extended Langlois' derivation to the case of two scalar fields minimally coupled to gravity.
The main challenge of the derivation lies in resolving the unknown coefficients $A_{\alpha\beta}$ and $B_\alpha$ in the quadratic ansatz for the function $S_{\scriptscriptstyle P}$ that generates the canonical transformation into the gauge-invariant variables.
The evolution of the gauge-invariant scalar perturbations are found to be coupled via the terms with $\Omega^2_{FF'}$ in the equations of motion \eqref{QF-EOM}, which may lead to novel phenomenology that cannot simply be inferred from that of single-field cosmological scenarios (see \cite{Frion:2025} for example).
We then delineated how the results for the scalar sector actually easily extend to the case with an arbitrary number of fields after a careful tracking of the scalar degrees of freedom of the system.
This constitutes the main result of this work, conveyed by the equations of motion for $n_f$ scalar gauge-invariant perturbations in \eqref{QF-EOM-multifield}.
With the aim of being pedagogically self-contained, we also recasted here the results in \cite{Langlois:1994ec} for the tensor sector, which are unaffected by the extension of the scalar sector (to second-order in linear perturbations).
Remarkably, the use of the Hamilton-Jacobi-\textit{like} procedure directly unveiled the gauge-invariant degrees of freedom of the system.
The generalisation presented in this work not only enables a Hamiltonian treatment of multi-field cosmological perturbations but also provides the foundation for a canonical quantisation of perturbations in models with multiple fields. As such, it lays the groundwork for future investigations into the quantum behaviour of coupled gauge-invariant perturbations in inflationary and alternative early-universe scenarios, especially in quantum cosmological frameworks.

\section{Acknowledgements}
We are grateful to Farshid Soltani, Fabio Mele, Carlo Rovelli and Francesca Vidotto for many helpful discussions throughout the drafting of this article.
We acknowledge that Western University is located on the traditional lands of the Anishinaabek, Haudenosaunee, Lūnaapéewak and Attawandaron peoples, on lands connected with the London Township and Sombra Treaties of 1796 and the Dish with One Spoon Covenant Wampum.


\bibliographystyle{myunsrt}
\bibliography{references}

\begin{thebibliography}{10}

\bibitem{Langlois:1994ec}
David Langlois.
\newblock Hamiltonian formalism and gauge invariance for linear perturbations in inflation.
\newblock {\em Class. Quant. Grav.}, 11(2):389--407, 1994.

\bibitem{Bardeen:1980kt}
James~M. Bardeen.
\newblock Gauge {Invariant} {Cosmological} {Perturbations}.
\newblock {\em Phys. Rev. D}, 22:1882--1905, 1980.

\bibitem{Kodama:1984ziu}
Hideo Kodama and Misao Sasaki.
\newblock Cosmological {Perturbation} {Theory}.
\newblock {\em Prog. Theor. Phys. Suppl.}, 78:1--166, 1984.

\bibitem{Mukhanov:1988}
V.~F. Mukhanov.
\newblock The quantum theory of gauge-invariant cosmological perturbations.
\newblock {\em Zhurnal Eksperimentalnoi i Teoreticheskoi Fiziki}, 94:1--11, July 1988.

\bibitem{Mukhanov:1990me}
V.~F. Mukhanov, H.~A. Feldman, and R.~H. Brandenberger.
\newblock Theory of cosmological perturbations.
\newblock {\em Phys. Rept.}, 215(5):203--333, 1992.

\bibitem{Mukhanov:2005book}
V.~Mukhanov.
\newblock {\em Physical {Foundations} of {Cosmology}}.
\newblock Cambridge University Press, December 2005.

\bibitem{Peter:2013woa}
Patrick Peter.
\newblock Cosmological {Perturbation} {Theory}.
\newblock {\em arXiv:1303.2509 [astro-ph.CO]}, March 2013.

\bibitem{Baumann:2022book}
Daniel Baumann.
\newblock {\em Cosmology}.
\newblock Cambridge University Press, 2022.

\bibitem{Planck:2018vyg}
N.~Aghanim, Y.~Akrami, M.~Ashdown, and {others}.
\newblock Planck 2018 results - {VI}. {Cosmological} parameters.
\newblock {\em Astron. Astrophys.}, 641:A6, September 2020.

\bibitem{SDSS:2005xqv}
Daniel~J. Eisenstein, Idit Zehavi, David~W. Hogg, and {others}.
\newblock Detection of the {Baryon} {Acoustic} {Peak} in the {Large}-{Scale} {Correlation} {Function} of {SDSS} {Luminous} {Red} {Galaxies}.
\newblock {\em Astrophys. J.}, 633(2):560--574, 2005.

\bibitem{Sasaki:1986hm}
Misao Sasaki.
\newblock Large {Scale} {Quantum} {Fluctuations} in the {Inflationary} {Universe}.
\newblock {\em Prog. Theor. Phys.}, 76(5):1036, 1986.

\bibitem{Langlois:2010xc}
D.~Langlois.
\newblock Inflation and {Cosmological} {Perturbations}.
\newblock In Georg Wolschin, editor, {\em Lect. {Notes} {Phys}.}, volume 800, pages 1--57. Springer Berlin Heidelberg, Berlin, Heidelberg, 2010.

\bibitem{Martin:2013tda}
Jerome Martin, Christophe Ringeval, and Vincent Vennin.
\newblock Encyclopædia {Inflationaris}.
\newblock {\em Phys. Dark Univ.}, 46:101653, October 2024.

\bibitem{Brandenberger:2009jq}
Robert~H. Brandenberger.
\newblock Alternatives to {Cosmological} {Inflation}.
\newblock {\em Int. J. Mod. Phys. Conf. Ser.}, 01:67--79, 2011.

\bibitem{Buchbinder:2007ad}
Evgeny~I. Buchbinder, Justin Khoury, and Burt~A. Ovrut.
\newblock New ekpyrotic cosmology.
\newblock {\em Phys. Rev. D}, 76(12):123503, 2007.

\bibitem{Brandenberger:2012zb}
Robert~H. Brandenberger.
\newblock The {Matter} {Bounce} {Alternative} to {Inflationary} {Cosmology}.
\newblock {\em ADS Harvard}, June 2012.

\bibitem{Wilson-Ewing:2012lmx}
Edward Wilson-Ewing.
\newblock The {Matter} {Bounce} {Scenario} in {Loop} {Quantum} {Cosmology}.
\newblock {\em JCAP}, 03:026, 2013.

\bibitem{Cai:2013kja}
Yi-Fu Cai, Evan McDonough, Francis Duplessis, and Robert Brandenberger.
\newblock Two {Field} {Matter} {Bounce} {Cosmology}.
\newblock {\em JCAP}, 10(10):024, October 2013.

\bibitem{Planck:2018jri}
Y.~Akrami, F.~Arroja, M.~Ashdown, and {others}.
\newblock Planck 2018 results - {X}. {Constraints} on inflation.
\newblock {\em Astron. Astrophys.}, 641:A10, September 2020.

\bibitem{Agullo:2020cvg}
Ivan Agullo, Dimitrios Kranas, and V.~Sreenath.
\newblock Large scale anomalies in the {CMB} and non-{Gaussianity} in bouncing cosmologies.
\newblock {\em Class. Quant. Grav.}, 38(6):065010, February 2021.

\bibitem{Agullo:2017eyh}
Ivan Agullo, Boris Bolliet, and V.~Sreenath.
\newblock Non-{Gaussianity} in loop quantum cosmology.
\newblock {\em Phys. Rev. D}, 97(6):066021, March 2018.

\bibitem{Nandi:2015ogk}
Debottam Nandi and S.~Shankaranarayanan.
\newblock Complete {Hamiltonian} analysis of cosmological perturbations at all orders.
\newblock {\em JCAP}, 06(06):038, June 2016.

\bibitem{Anderegg:1994xq}
S.~Anderegg and V.~Mukhanov.
\newblock Path {Integral} {Quantization} of {Cosmological} {Perturbations}.
\newblock {\em Phys. Lett. B}, 331(1-2):30--38, 1994.

\bibitem{Agullo:2016tjh}
Ivan Agullo and Parampreet Singh.
\newblock Loop {Quantum} {Cosmology}: {A} brief review.
\newblock {\em World Scientific}, 2017.

\bibitem{Ashtekar:2011ni}
Abhay Ashtekar and Parampreet Singh.
\newblock Loop {Quantum} {Cosmology}: {A} {Status} {Report}.
\newblock {\em Class. Quant. Grav.}, 28(21):213001, 2011.

\bibitem{Bojowald:2008zzb}
Martin Bojowald.
\newblock Loop {Quantum} {Cosmology}.
\newblock {\em Living Rev. Rel.}, 11(1):4, 2008.

\bibitem{Frion:2025}
{Emmanuel Frion}, {Mateo Pascual}, and {Francesca Vidotto}.
\newblock Quasi-dust {Ekpyrotic} {LQC} bounce scenario.
\newblock {\em In preparation}, April 2025.

\bibitem{Domenech:2017ems}
Guillem Domènech and Misao Sasaki.
\newblock Hamiltonian approach to second order gauge invariant cosmological perturbations.
\newblock {\em Phys. Rev. D}, 97(2):023521, January 2018.

\bibitem{wald_general_1984}
Robert~M Wald.
\newblock {\em General relativity}.
\newblock University of Chicago Press, Chicago ; London, 1984.

\bibitem{dirac_lectures_1964}
P.A.M. Dirac.
\newblock {\em Lectures on quantum mechanics}.
\newblock Belfer Graduate School of Science, Yeshiva University, New York, 1964.

\bibitem{Mele:2021gzx}
Fabio~M. Mele and Johannes Münch.
\newblock The {Physical} {Relevance} of the {Fiducial} {Cell} in {Loop} {Quantum} {Cosmology}.
\newblock {\em Phys. Rev. D}, 108(10):106004, November 2023.

\bibitem{Mele:2022quh}
Fabio~M. Mele and Johannes Münch.
\newblock On the {Role} of {Fiducial} {Structures} in {Minisuperspace} {Reduction} and {Quantum} {Fluctuations} in {LQC}.
\newblock {\em Class. Quant. Grav.}, 41(24):245003, November 2024.

\bibitem{Artigas:2021zdk}
Danilo Artigas, Julien Grain, and Vincent Vennin.
\newblock Hamiltonian formalism for cosmological perturbations: the separate-universe approach.
\newblock {\em JCAP}, 02(02):001, February 2022.

\bibitem{goldberg_hamiltonian_1991}
Joshua Goldberg, Ezra~T. Newman, and Carlo Rovelli.
\newblock On {Hamiltonian} systems with first‐class constraints.
\newblock {\em Journal of Mathematical Physics}, 32(10):2739--2743, October 1991.

\bibitem{goldstein_classical_1950}
Herbert Goldstein.
\newblock {\em Classical {Mechanics}}.
\newblock Addison-Wesley, 1950.

\bibitem{Gong:2016qmq}
Jinn-Ouk Gong.
\newblock Multi-field inflation and cosmological perturbations.
\newblock {\em Int. J. Mod. Phys. D}, 26(01):1740003, August 2016.

\end{thebibliography}

\end{document}